\documentclass[journal=jctcce,manuscript=article,etalmode=truncate]{achemso}
\setkeys{acs}{etalmode=truncate,maxauthors=0,articletitle=true}

\usepackage[version=3]{mhchem}
\usepackage[T1]{fontenc}

\usepackage{amsmath,amsfonts,amssymb,mathtools,mymath}
\usepackage[algo2e,ruled]{algorithm2e}
\usepackage{hyperref,cleveref}
\crefname{equation}{}{}

\usepackage{subfig}
\newif\ifexporttikz
\exporttikztrue 
\ifexporttikz
  \usepackage{tikz}
  \usepackage{pgfplots}
  \pgfplotsset{
    compat=newest,
    table/header=false,
    title style={font=\small},
    tick label style={font=\scriptsize},
    label style={font=\scriptsize},
    legend style={font=\scriptsize},
    legend cell align=left
  }
  \usepgfplotslibrary{external}
\else
  \usepackage{tikzexternal}
  \tikzsetexternalprefix{fig/}
  
\fi
\newcommand{\figname}[1]{\tikzsetnextfilename{#1}}
\newcommand{\fignames}[1]{\tikzsetfigurename{#1}}
\newcommand{\datfile}[1]{fig/#1.dat}

\usepackage{tabularx,multirow,booktabs}
\newcolumntype{L}{>{\raggedright\arraybackslash}X}
\newcolumntype{C}{>{\centering\arraybackslash}X}
\newcolumntype{R}{>{\raggedleft\arraybackslash}X}

\newcommand\bZero{\mathbf{0}}
\newcommand\bMKh{\widehat{\bM\bK}}
\newcommand\eV{\mathrm{eV}}

\title{A Model Order Reduction Algorithm for Estimating the Absorption Spectrum}

\author{Roel Van Beeumen}
\email{rvanbeeumen@lbl.gov}
\affiliation{Computational Research Division, Lawrence Berkeley National Laboratory, Berkeley, CA}
\author{David B. Williams-Young}
\email{dbwy@uw.edu}
\affiliation{Department of Chemistry, University of Washington, Seattle, WA}
\author{Joseph M. Kasper}
\email{jkasper2@uw.edu}
\affiliation{Department of Chemistry, University of Washington, Seattle, WA}
\author{Chao~Yang}
\email{cyang@lbl.gov}
\affiliation{Computational Research Division, Lawrence Berkeley National Laboratory, Berkeley, CA}
\author{Esmond G. Ng}
\email{egng@lbl.gov}
\affiliation{Computational Research Division, Lawrence Berkeley National Laboratory, Berkeley, CA}
\author{Xiaosong Li}
\email{xsli@uw.edu}
\affiliation{Department of Chemistry, University of Washington, Seattle, WA}

\begin{document}


\def\PARn{2}
\def\PARk{1}
\def\PARX{4.5}
\def\PARY{4.9}
\def\PARdx{0.2}

\begin{figure}
\centering
\figname{abstract_graphic}%
\fbox{\begin{tikzpicture}
\begin{axis}[%
 axis equal,%
 width=3.6in,%
 height=2.1in,%
 axis x line=bottom,%
 axis y line=left,%
 xlabel={Energy (eV)},%
 ylabel={Absorption Spectrum},%
 xmin=0,xmax=10,%
 ymin=0,ymax=5,%
 xtick={-1},%
 ytick={-1},%
]
\addplot[mark=,thick,blue] table[x index=0,y index=1]{\datfile{abstract_graphic_refsol}};
\addplot[ycomb,gray] table[x index=0,y index=1]{\datfile{abstract_graphic_oscstr}};
\draw[fill=white!95!black] ({\PARX-4*\PARdx},{\PARY}) rectangle ({\PARX-4*\PARdx-\PARk},{\PARY-\PARk});
\draw[fill=white!95!black] ({\PARX},{\PARY}) rectangle ({\PARX+\PARn},{\PARY-\PARk});
\draw[fill=white!95!black] ({\PARX+\PARn+\PARdx},{\PARY}) rectangle ({\PARX+\PARn+\PARdx+\PARn},{\PARY-\PARn});
\draw[fill=white!95!black] ({\PARX+2*\PARn+2*\PARdx},{\PARY}) rectangle ({\PARX+2*\PARn+2*\PARdx+\PARk},{\PARY-\PARn});
\draw ({\PARX-4*\PARdx-\PARk/2},{\PARY-\PARk/2}) node {\scriptsize$\bHh$};
\draw ({\PARX-2*\PARdx},{\PARY-\PARk/2}) node {\scriptsize$=$};
\draw ({\PARX+\PARn/2},{\PARY-\PARk/2}) node {\scriptsize~$\bV\T$};
\draw ({\PARX+\PARn+\PARdx+\PARn/2},{\PARY-\PARn/2}) node {\scriptsize$\bH$};
\draw ({\PARX+2*\PARn+2*\PARdx+\PARk/2},{\PARY-\PARn/2}) node {\scriptsize$\bV$};
\end{axis}
\end{tikzpicture}%
}
\end{figure}

\begin{abstract}
The \emph{ab initio} description of the spectral interior of the absorption
spectrum poses both a theoretical and computational challenge for modern
electronic structure theory. Due to the often spectrally dense character of
this domain in the quantum propagator's eigenspectrum for medium-to-large sized
systems, traditional approaches based on the partial diagonalization of the
propagator often encounter oscillatory and stagnating convergence.
Electronic structure methods which solve the molecular response problem
through the solution of spectrally shifted linear systems, such as the complex
polarization propagator, offer an alternative approach which is agnostic to the
underlying spectral density or domain location. This generality comes at a
seemingly high computational cost associated with solving a large linear system
for each spectral shift in some discretization of the spectral domain of
interest. In this work, we present a novel, adaptive solution to this high
computational overhead based on model order reduction techniques via
interpolation. Model order reduction reduces the computational complexity of
mathematical models and is ubiquitous in the simulation of dynamical systems
and control theory. The efficiency and effectiveness of the proposed algorithm
in the \emph{ab initio} prediction of X-Ray absorption spectra is demonstrated
using a test set of challenging water clusters which are spectrally dense in
the neighborhood of the oxygen $K$-edge. Based on a single, user defined
tolerance we automatically determine the order of the reduced models and
approximate the absorption spectrum up to the given tolerance. We also
illustrate that, for the systems studied, the automatically determined model order
increases logarithmically with the problem dimension, compared to a linear increase of the
number of eigenvalues within the energy window. Furthermore, we observed that
the computational cost of the proposed algorithm only scales quadratically with
respect to the problem dimension.
\end{abstract}

\section{Introduction}
\label{sec:intro}

With recent advances in  laser light source technology, X-ray
absorption spectroscopy (XAS) has become an important probative tool in
chemical physics.\cite{Stohr13_book} The ability of XAS to simultaneously
characterize both the electronic and geometrical structure of chemical systems
has made it indispensable in the fields of catalysis and
photophysics.\cite{Koch87_519,Chasse12_4870,Solomon95_2259,Hodgson00_5775,Hessler01_262}
However, despite the capability of XAS to obtain a wealth of chemically
relevant information, the complexity of experimentally obtained XAS spectra
often requires a theoretical supplement to obtain a meaningful interpretation
of the query phenomenon.\cite{Li16_639,Li16_JA2} Thus, the ability to properly
describe the high-energy electronic excitations of molecular systems
theoretically is critical in modern electronic structure theory.

In light of its importance in physical chemistry, the prediction of XAS properties poses an interesting challenge for traditional electronic structure methods. 
This challenge is rooted in the fact that the X-Ray region is buried deep within the eigenspectrum of the Hamiltonian and is often spectrally dense. 
For example, in near edge X-Ray absorption fine structure (NEXAFS) spectroscopy, the spectrum consists of many excited states that correspond to excitations of core electrons to diffuse quasi bound levels. 
Thus, as system sizes increase, the number of states in the given energy region increases dramatically. 
Further, it is important to note that, because very large basis sets are often required to properly describe the rather diffuse nature of these excited states, the increase in complexity leads to poor scaling with system size.

Many electronic structure methods have been extended to the
description of high-energy, X-ray electronic excitations in recent years. In
the time domain, real-time density functional theory\cite{Li05_233,Li07_199,Li11_184102} has been shown to
excellently reproduce the X-ray \emph{K}-edge for molecules within relatively
short simulation times.\cite{Govind12_3284,Lopata16_3741} For large systems,
however, time-domain methods have difficulty taking full advantage of
concurrency on modern computing architectures, and are thus not yet a
sustainable avenue in routine theoretical inquiry of these phenomena. In contrast,
frequency domain approaches are often favored in these types of calculations as
they may be cast as computationally scalable linear algebra problems which are
well suited for massive concurrency. Frequency domain approaches to treat
electronic excitations may be separated into two categories which obtain equivalent information:
methods which aim to obtain a spectral decomposition of the quantum propagator, i.e., eigenproblem based methods, and methods which solve the response problem directly through the solution of linear systems of equations. 

Recasting electronic structure methods into eigenproblems has long been the de facto standard frequency domain method for electronically excited states. Through knowledge of the poles (eigenroots) of the quantum propagator, one has direct access to information regarding the electronic excitations (resonances) of the molecular system. In addition, such a spectral decomposition may be used to treat off-resonant perturbations through interpolation schemes known as sum-over-states expressions\cite{Yeager84_33}.
Much work has gone into the development of these methods in both wave function theory, such as those based on
the 
coupled-cluster (CC)\cite{olje1988,Monkhorst77_421,Jorgensen90_3333,Bartlett93_7029,Bartlett93_414} 
and algebraic diagrammatic construction (ADC)\cite{Dreuw14_4583,Dreuw14_1900} 
expansions of the many-body wave function,
and self-consistent field theory, such as the linear response time-dependent Hartree--Fock (TD-HF)\cite{Hattig98_1,Ring_book,Jorgensen_book,Rowe68_153} and density functional theory
(TD-DFT)\cite{Casida95_Book,HeadGordon05_4009}. These methods have been shown to accurately predict and reproduce both low-\cite{Ruud12_543,Bartlett09_Book} and high-energy\cite{Li11_3540,Li15_2994,Li15_4146,DeSimone03_115,Neese07_2783,Asmuruf10_12024,Govind12_3284} electronic excitations in molecular systems.  Despite
their accuracy, however, eigenproblem based methods possess an inherent challenge in the
description of high-energy excited states when the eigenroots of interest
are buried deep in the eigenspectrum.  Traditional methods used to partially
diagonalize the propagator, such as the
block-Davidson method\cite{Davidson75_87,Scott86_817,Morgan92_287}, are designed to converge to the extreme ends of the
eigenspectrum with no built-in mechanism to establish the spectrum's interior. 
Several approaches have been described to overcome this problem~\cite{zuev_etal2015}, including
energy specific\cite{Li11_3540,Li15_4146} and
restricted energy window methods\cite{DeSimone03_115,Neese07_2783,Asmuruf10_12024} when the eigenroots of interest are well-separated. Further, in spectrally dense regions of the propagator's eigenspectrum, iterative diagonalization algorithms require the resolution of many more roots than is often practical to ensure smooth convergence.

Methods which solve the response problem through the solutions of linear systems offer an attractive alternative to eigenproblem based approaches in the description of high-energy excitations because they have an intrinsic
mechanism to probe the interior of the energy spectrum. In these methods, the
probing frequency of the applied perturbation is a chosen parameter.\cite{Hattig98_1,Ruud12_543} Thus,
the interior of the spectrum is easily probed through a number of solutions
of linear system of equations in the desired frequency domain. This simplicity 
does, however, come at a seemly significant computational cost compared to eigenproblem based methods.
While eigenproblems are able to directly obtain many poles of the eigenspectrum
simultaneously, one must solve the linear problem many times
over some discretization of the frequency domain to obtain similar results.
In general, this discretization must be quite dense to achieve a reasonable 
accuracy and thus can be more expensive than their eigenproblem based counterparts. Approaches using linear systems and based on the complex
polarization propagator (CPP), such as CPP-CC\cite{Norman12_1616,Norman13_124311,Coriani13_211102} and CPP-SCF,\cite{Rubio_Book,Ruud12_543,Yeager84_33,Oddershede01_JCP}
have been shown to be successful in the description of both high\cite{Norman16_1991,Norman12_022507,Norman10_5096,Agren06_143001,Norman16_13591} and low\cite{Mathieu15_21866} energy properties of molecular systems and have been extended to relativistic Hamiltonians as well\cite{Norman10_064105}.

In this work, we introduce a general framework for the prediction of spectrally interior molecular response properties based on model order reduction (MOR) via interpolation. MOR techniques have been successfully applied in different fields of computation science and engineering, where it reduces the computational complexity of mathematical models in numerical simulations. Examples include structural dynamics, sound and vibration analysis, and control theory \cite{Antoulas2005}. The MOR algorithm proposed in this paper aims to overcome the large computational overhead associated with the spectral discretization required by linear system based methods while maintaining the accuracy associated with eigenproblem based methods. Further, the proposed algorithm will be shown to allow for the massively scalable parallelism that is well suited for modern computing architectures.

\section{Linear response and absorption spectrum}
\label{sec:lras}

In the semi-classical theory of molecular light-matter interaction within the electric dipole approximation, the isotropic absorption cross section for the interaction with plane-polarized light, $\sigma(\omega)$, at a particular perturbing frequency, $\omega$, is proportional to the trace of the dynamic polarizability tensor, $\bfalpha(\omega)$,
\begin{equation}
  \sigma(\omega) \propto \omega \imag\left(\trace\left[ \bfalpha(\omegat) \right]\right), \qquad \omegat = \omega + i\eta,
  \label{eq:abs-spectrum}
\end{equation}
where $\eta > 0$ is a small damping parameter to ensure the convergence of $\bfalpha$ in the spectral neighborhoods of resonant perturbations. Within the linear response regime of the first-order polarization propagator approximation (FOPPA)\cite{Yeager84_33}, the dynamic polarizability tensor may be written as
\begin{equation}
  \bfalpha(\omegat) = \bfd\T \bG\inv(\omegat) \bfd, \qquad 
  \bfd = \begin{bmatrix}
    \bfd_x & \bfd_y & \bfd_z \\
    \bfd_x & \bfd_y & \bfd_z
  \end{bmatrix}.
  \label{eq:polar-tensor}
\end{equation}
Here, $\{ \bfd_i\text{ } \vert\text{ } i \in \{ x,y,z \} \}$ is the set of dipole operators expressed in the molecular orbital (MO) basis, and $\bG(\omegat)$ is the first-order polarization propagator. In the following algorithm developments, we restrict the discussion to the FOPPA using a Hartree--Fock reference (TD-HF), although the algorithm presented is completely general to any choice of propagator or reference. Within TD-HF, $\bG(\omegat)$ may be written as
\begin{equation}
  \bG(\omegat) = \bH - \omegat\bS,
  \label{eq:defG}
\end{equation}
where
\begin{equation}
  \bH = \begin{bmatrix}
    \bA & \bB \\
    \bB & \bA
  \end{bmatrix}, \qquad
  \bS = \begin{bmatrix*}[r]
    \bI & \bZero \\
    \bZero & -\bI
  \end{bmatrix*},
  \label{eq:defH}
\end{equation}
with $\bS = \bS\T = \bS\inv$ and
\begin{align}
  A_{ai,bj} &= \delta_{ij}\delta_{ab}\left( \epsilon_a - \epsilon_i \right) + (ai \vert \vert bj ),\\
  B_{ai,bj} &= (ab \vert \vert ij ).
\end{align}
Here, we have adopted the convention of denoting occupied MOs with indices $i,j,k,\ldots$ and virtual (unoccupied) MOs with indices $a,b,c,\ldots$ $\{\epsilon_p\}$ is taken to be the set of canonical Hartree--Fock MO eigenenergies and the moieties $(\cdot \vert\vert \cdot)$ are the MO basis representation of the anti-symmetrized electron-repulsion integrals in Mulliken notation. In this work, we restrict our treatment to the use of strictly real MOs to allow for further simplification of the working expressions.

In order to study the spectrum of the pencil $(\bH,\bS)$ let
\begin{equation}
  \bOmega = \bS\inv\bH = \begin{bmatrix*}[r]
    \bA & \bB \\
    -\bB & -\bA
  \end{bmatrix*}.
  \label{eq:defOmega}
\end{equation}
Although the matrix $\bOmega$ is non-symmetric, it has a number of special properties \cite{olje1988,beme1998,bali2012}. 
If $\bH$ is positive definite, it may be shown that $\bOmega$ possesses a structured eigendecomposition \cite{Jorgensen_book,SJYDL2016}, i.e.,
\begin{equation}
  \begin{bmatrix*}[r] \bA & \bB \\ -\bB & -\bA \end{bmatrix*}
  = \begin{bmatrix*}[r] \bU & \bV \\ \bV & \bU \end{bmatrix*}
    \begin{bmatrix*}[r] \bLambda & \bZero \\ \bZero & -\bLambda \end{bmatrix*}
    \begin{bmatrix*}[r] \bU & -\bV \\ -\bV & \bU \end{bmatrix*}\T
\end{equation}
where $\bLambda = \diag(\lambda_1,\dotsc,\lambda_n)$ consists of strictly positive eigenvalues, 
and the eigenvectors are normalized with respect to the metric $\bS$,
\begin{equation}
  \begin{bmatrix*}[r] \bU & -\bV \\ -\bV & \bU \end{bmatrix*}\T
  \begin{bmatrix*}[r] \bU &  \bV \\  \bV & \bU \end{bmatrix*} = \bI.
  \label{eq:eigvec-normalization}
\end{equation}

As $\bH$ is taken to be real in this work, it possesses additional properties that may be exploited in the development of efficient algorithms for estimating the absorption spectrum of the target system. In particular, we may apply the following similarity transformation
\begin{equation}
  \bT = \frac{1}{\sqrt{2}} \begin{bmatrix*}[r]
     \bI & \bI \\
    -\bI & \bI
  \end{bmatrix*}, \qquad \bT\inv = \bT\T,
\label{eq:defT}
\end{equation}
to $\bG(\omegat)$, yielding
\begin{equation}
  \bT\T \bG(\omegat) \bT = \begin{bmatrix}
    \bK & \bZero \\
    \bZero & \bM
  \end{bmatrix} - \omegat \begin{bmatrix}
    \bZero & \bI \\
    \bI & \bZero
  \end{bmatrix},
  \label{eq:}
\end{equation}
where
\begin{align}
  \bM &\equiv \bA + \bB, \label{eq:defM} \\
  \bK &\equiv \bA - \bB, \label{eq:defK}
\end{align}
which are, in most cases, positive definite. In this case, the polarizability tensor may be reformulated as
\begin{equation}
  \bfalpha(\omegat) = \bfdt\T \bGt\inv(\omegat) \bfdt, \qquad \bfdt = \begin{bmatrix}
    \bfd_x & \bfd_y & \bfd_z
  \end{bmatrix},
  \label{eq:polar-tensor-MK}
\end{equation}
where
\begin{equation}
  \bGt(\omegat) = \bM\bK - \omegat^2\bI.
  \label{eq:defGt}
\end{equation}
Note that the dimension of $\bGt(\omegat)$ is only half the dimension of $\bG(\omegat)$. Furthermore, it can be shown that
\begin{align}
 \bM &= (\bX - \bY) \bLambda (\bX - \bY)\T, \\
 \bK &= (\bX + \bY) \bLambda (\bX + \bY)\T,
\end{align}
and
\begin{equation}
  (\bX - \bY)\T (\bX + \bY) = \bI,
\end{equation}
such that the eigenvalues $\pm\bLambda$ may be computed by
\begin{equation}
  \bM\bK = (\bX - \bY) \bLambda^2 (\bX + \bY)\T.
\end{equation}
Remark that by making use of $\bM\bK$, the dimension of the eigenvalue problem is also reduced by a factor of 2.\cite{Haser93_1262,Frisch98_8218}

\section{Model order reduction of linear dynamical systems}
\label{lds:dyn-sys}

In this section, we briefly review the theory of model order reduction for linear dynamical systems. The next section will examine its connection to the computation of the absorption spectrum within the FOPPA.

\subsection{Linear dynamical systems}
\label{lds:mimo}

We consider the linear multiple-input multiple-output (MIMO) system
\begin{equation}
  \bSigma = \left\{\begin{aligned}
    \left( \bH - s\bS \right) \bfx(s) &= \bfb\,u(s) \\
                              \bfy(s) &= \bfc\T \bfx(s)
  \end{aligned}\right.,
  \label{eq:siso}
\end{equation}
where $s$ is a derivative or shift operator, $\bH \inRR{n}$ and $\bS \inRR{n}$ are the system matrices, $\bfb \inR[n][m]$, and $\bfc \inR[n][p]$. We call $n$ the dimension  (order) of the system $\bSigma$, $\bfx \inR[n][m]$ the state vector, $u \inR$ the input, and $\bfy \inR[p][m]$ the output \cite{Antoulas2005}. Note that the system $\bSigma$ is completely characterized by the quadruple $(\bH,\bS,\bfb,\bfc)$.

The transfer function, $\bfgamma(s)$, of $\bSigma$ is defined as
\begin{equation}
  \bfgamma(s) = \bfc\T \left( \bH - s\bS \right)\inv \bfb,
  \label{eq:siso-tf}
\end{equation}
and describes the relation between the input and output of $\bSigma$, i.e., $\bfy(s) = \bfgamma(s) u(s)$. For the remainder, we will assume that $n \gg 1$, $m \ll n$, $p \ll n$, and $u(s) \equiv 1$ for all $s$.

\subsection{State space transformation}
\label{lds:ss-transf}

In some cases, it might be more advantageous to describe the system from a different point of view as the original one. In these cases, we may perform a non-singular state transformation $\bT$, i.e., $\det(\bT) \neq 0$, yielding the transformed state
\begin{equation}
  \bfxt = \bT\inv \bfx,
  \label{eq:defxtil}
\end{equation}
of the transformed system
\begin{equation}
  \bSigmat = \left\{\begin{aligned}
    \left( \bHt - s\bSt \right) \bfxt(s) &= \bfbt\,u(s) \\
                                 \bfy(s) &= \bfct\T \bfxt(s)
  \end{aligned}\right.,
\end{equation}
where $\bHt = \bT\inv \bH \bT$, $\bSt = \bT\inv \bS \bT$, $\bfbt = \bT\inv \bfb$, and $\bfct\T = \bfc\T \bT$. Remark that $\bSigma$ and $\bSigmat$ admit the same transfer function as well as the same output. Therefore, we call the systems $\bSigma$ and $\bSigmat$ equivalent.

\subsection{Reduced order models}
\label{lds:mor}

The evaluation of the transfer function of a system $\bSigma$ requires a linear system solve for every value of $s$. In cases where the system dimension $n$ is large and a high resolution is required, i.e., a high number of values of $s$, the evaluation of the transfer function is very expensive. In this work, we examine the effectiveness of model order reduction (MOR) techniques to circumvent this expense. MOR for linear dynamical systems is a technique that approximates a system $\bSigma$ by another system $\bSigmah$ of the same form but of a much lower dimension (order) $k \ll n$. Consequently, evaluating the transfer function of $\bSigmah$ is relatively inexpensive as it only involves linear system solves of dimension $k$ instead of linear system solves of dimension $n$ for $\bSigma$.

Let the system $\bSigma$ be given by \eqref{eq:siso} and define a non-singular matrix $\bV \inR[n][k]$ with orthonormal columns, i.e., $\bV\T \bV = \bI$. Then, a reduced order model $\bSigmah$ can be constructed by applying a Galerkin projection $\bP = \bV\bV\T$ onto $\bSigma$, yielding
\begin{equation}
  \bSigmah = \left\{\begin{aligned}
    \left( \bHh - s\bSh \right) \bfxh(s) &= \bfbh\,u(s) \\
                                \bfyh(s) &= \bfch\T \bfxh(s)
  \end{aligned}\right.,
\end{equation}
where $\bHh = \bV\T \bH \bV$, $\bSh = \bV\T \bS \bV$, $\bfbh = \bV\T \bfb$, and $\bfch\T = \bfc\T \bV$. Note that the length of the state vector $\bfxh$ and the dimension of $\bSigmah$ are only $k \ll n$. The purpose of MOR is to construct a $\bV$ such that the transfer function of $\bSigmah$ approximates very well the one of $\bSigma$,
\begin{equation}
  \bfgamma_\bSigma(s) \approx \bfgamma_{\bSigmah}(s),
  \label{eq:tf-match}
\end{equation}
for all query $s$.

\subsection{Model order reduction via moment matching}
\label{lds:mm}

One way to construct a matrix $\bV$ such that \eqref{eq:tf-match} holds is by examining the concepts of moments and moment matching\cite{Antoulas2005}. Let the transfer function $\bfgamma$ of $\bSigma$ be given by \eqref{eq:siso-tf}. Then the $\ell$th moment of $\bfgamma$ around the point $s = s_\star$ is defined as the $\ell$th derivative of $\bfgamma$ evaluated at $s_\star$, i.e.,
\begin{equation}
  \bfm_\ell(s_\star) := (-1)^\ell \left.\frac{d^\ell}{ds^\ell} \bfgamma(s) \right|_{s=s_\star},
  \label{eq:moment}
\end{equation}
for $\ell \geq 0$. Consequently, since $\bfgamma(s) = \bfc\T \left( \bH - s\bS \right)\inv \bfb$, the moments at $s_\star$ are
$$
  \bfm_\ell(s_\star) = \bfc\T \left( \bH - s_\star \bS \right)^{-(\ell+1)} \bfb, \qquad \ell > 0.
$$
Note also that the moments determine the coefficients of the Taylor series expansion of the transfer function $\bfgamma$ in the neighborhood of $s_\star$
\begin{equation}
  \bfgamma(s) = \bfm_0(s_\star) + \bfm_1(s_\star) \frac{s - s_\star}{1!} + \bfm_2(s_\star) \frac{(s - s_\star)^2}{2!} + \cdots
\end{equation}

Model order reduction via moment matching consists of constructing a subspace $\bV \inR[n][km]$ such that the original and reduced order model match moments
\begin{equation}
  \bfm_{i_j}(s_j) = \bfmh_{i_j}(s_j), \qquad j = 1,\ldots,k.
  \label{eq:mor-mm}
\end{equation}
If all moments to be matched are chosen at zero, i.e., $s_j = 0$ for $j = 1,2,\ldots,k$, the corresponding model is known as a Pad\'{e} approximation. In the general case, the problem \eqref{eq:mor-mm} is known as rational interpolation and can be solved by choosing the projection matrix $\bV$ such that
\begin{equation}
  \bV = \spn \Big[ \left( \bH - s_1\bS \right)\inv \bfb \quad
                   \left( \bH - s_2\bS \right)\inv \bfb \quad
                    \cdots \quad
                   \left( \bH - s_k\bS \right)\inv \bfb \Big].
  \label{eq:defV-app}
\end{equation}
It can be shown that the matrix $\bV$ defined in \eqref{eq:defV-app} spans a rational Krylov subspace and matches all the $0$th moments at $s_j$. For more information about the connections between moment matching and rational interpolation, we refer the interested reader to Section~11 of Antoulas' model order reduction book \cite{Antoulas2005}.

\section{Estimating absorption spectrum without explicitly computing eigenvalues and eigenvectors}
\label{sec:est}

The most straightforward way to evaluate the absorption spectrum is to compute eigenvalues and the corresponding eigenvectors of $(\bH,\bS)$. However, as we indicated earlier, when the dimension of $\bH$ and $\bS$ becomes large (spectrally dense), this approach can be prohibitively expensive (complicated). 

It has been shown\cite{brabec_etal2015} that a special $\bK$-inner product
Lanczos algorithm can be used to provide a good approximation to the overall
structure of the absorption spectrum without explicitly computing the
eigenvalues and eigenvectors of $(\bH,\bS)$.  In particular, the Lanczos
algorithm can reveal major absorption peaks in the low frequency region of the
spectrum without too many iterations. However, the algorithm gives limited
resolution of the absorption spectrum in the spectral interior as the
Krylov subspace constructed by the Lanczos iteration contains little spectral
information associated with interior eigenvalues of $(\bH,\bS)$.

We now propose an alternative way to evaluate the absorption spectrum without explicitly computing the eigenvalues and eigenvectors of $(\bH,\bS)$. This scheme focuses on approximating the dynamic polarizability tensor $\bfalpha(\omegat)$ defined in \eqref{eq:polar-tensor} and the absorption spectrum $\sigma(\omega)$ defined in \eqref{eq:abs-spectrum} within a specific energy window directly.

Firstly, observe that the dynamic polarizability tensor \eqref{eq:polar-tensor} may be viewed simply as the expectation value of the inverse of $\bH -\omegat \bS$. Hence, the evaluation of $\bfalpha(\omegat)$ may be recast into a problem of solving linear equations, i.e., for a specific frequency $\omega$, we can directly evaluate the absorption spectrum \eqref{eq:abs-spectrum} as follows
\begin{equation}
  \sigma(\omega) \propto \omega \imag\left(\trace\left[ \bfd\T \bfx(\omegat) \right]\right),
\end{equation}
where $\bfx$ is the solution of the linear system
\begin{equation}
  \left( \bH - \omegat\bS \right) \bfx(\omegat) = \bfd.
  \label{eq:lin-sys}
\end{equation}

Secondly, the dynamic polarizability tensor \eqref{eq:polar-tensor} may also be viewed as the transfer function, i.e., the relation between input and output, of the linear dynamical system (see \cref{lds:mimo})
\begin{equation}
  \left\{\begin{aligned}
    \left( \bH - \omegat\bS \right) \bfx(\omegat) &= \bfd \\
                                     \bfy(\omega) &= \bfd\T \bfx(\omegat)
  \end{aligned}\right..
  \label{eq:MIMO}
\end{equation}
Consequently, the absorption spectrum can directly be obtained from the output variable $\bfy$
\begin{equation}
  \sigma(\omega) \propto \omega \imag\left(\trace\left[ \bfy(\omega) \right]\right).
\end{equation}
In order to evaluate the output $\bfy$ of system \eqref{eq:MIMO} for a given frequency, we again need to solve a linear system of the form \eqref{eq:lin-sys}.

Finally, by exploiting the block structure of $\bH$ and performing a state space transformation with \eqref{eq:defT} (see \cref{lds:ss-transf}), we obtain an equivalent linear dynamical system for \eqref{eq:MIMO}, but with a halved order,
\begin{equation}
  \left\{\begin{aligned}
    \left( \bM\bK - \omegat^2\bI \right) \bfxt(\omegat) &= \bfdt \\
                                           \bfy(\omega) &= 2\,\bfdt\T \bK\,\bfxt(\omegat)
  \end{aligned}\right.,
  \label{eq:MIMO-MK}
\end{equation}
such that we obtain the following, compact expressions for the dynamic polarizability tensor
\begin{equation}
  \bfalpha(\omegat) = 2\,\bfdt\T \bK \left( \bM\bK - \omegat^2\bI \right)\inv \bfdt,
  \label{eq:polar-tensor-lin-MK}
\end{equation}
and the absorption spectrum
\begin{equation}
  \sigma(\omega) \propto \omega \imag\left(\trace\left[ \bfdt\T \bK \left( \bM\bK - \omegat^2\bI \right)\inv \bfdt \right]\right).
  \label{eq:abs-spectrum-lin-MK}
\end{equation}
Note that the dimension of the linear systems to be solved in \eqref{eq:abs-spectrum-lin-MK} is only half of the dimension of the linear system shown in \eqref{eq:lin-sys}.

Clearly, we cannot afford to evaluate $\sigma(\omega)$ for all $\omega$'s of interest. However, this connection to linear dynamical systems allows us to employ MOR techniques (see \cref{lds:mor}) to reduce the number of $\sigma(\omega)$ evaluations in the full dimension. More precisely, we construct a function $\hat{\sigma}(\omega)$ that approximates $\sigma(\omega)$ within a specific energy window $[\omega_\mathrm{min},\omega_\mathrm{max}]$, but is much cheaper to evaluate.  The construction of such an approximate function only requires solving a few linear systems of the form \eqref{eq:lin-sys} or \eqref{eq:MIMO-MK} at a few selected frequencies $\tau_j$, $j = 1,2,\ldots,k$. The solutions of these linear systems are then used to construct a reduced order model which interpolates the full dynamic polarizability at $\tau_j$, and provides an approximation to the dynamic polarizability tensor \eqref{eq:polar-tensor} at other frequencies within the predefined energy window. When $k$ is small, both the construction and the evaluation of the reduced order model is significantly lower than other approaches that are either based on solving an eigenvalue problem or \eqref{eq:lin-sys} at many different frequencies.

\section{Interpolation based algorithms}
\label{sec:mor}

Let the dimension of the matrix $\bH$ defined in \eqref{eq:defH} be $2n \times 2n$. The dimension of the lower dimensional matrix $\bHh$ that we construct for the reduced order model is $3k \times 3k$, where $k \ll n$. One way to construct such a matrix is to first construct a subspace spanned by orthonormal columns of a matrix $\bV \inR[2n][3k]$ and then project $\bH$ onto such a subspace $\bV$, i.e.,
\begin{equation}
  \bHh = \bV\T \bH \bV.
\end{equation}
If we also let $\bSh = \bV\T \bS \bV$ and $\bfdh = \bV\T \bfd$, then the absorption spectrum can be approximated by
\begin{equation}
  \sigmah(\omega) \propto \omega \imag\left(\trace\left[ \bfdh\T \left( \bHh - \omegat\bSh \right)\inv \bfdh \right]\right).
  \label{eq:abs-spectrum-k}
\end{equation}

Clearly, the choice of the subspace $\bV$ is crucial in maintaining the 
fidelity of the reduced order model.  The subspace we use to construct the reduced order model takes the form
\begin{equation}
  \bV = \spn \Big[ \left( \bH - \tau_1\bS \right)\inv \bfd \quad \left( \bH - \tau_2\bS \right)\inv \bfd \quad \cdots \quad \left( \bH - \tau_k\bS \right)\inv \bfd \Big],
  \label{eq:defV}
\end{equation}
where $\tau_j$, $j = 1,2,\ldots,k$, are the interpolation frequencies carefully 
chosen within the energy window of interest to ensure that
\begin{equation}
  \sigma(\omega) \approx \sigmah(\omega),
\end{equation}
for all $\omega$ in the energy window of interest. It follows from the way $\bV$ is constructed in \eqref{eq:defV} that $\bfalphah$ interpolates $\bfalpha$ at the interpolation frequencies, i.e.,
\begin{equation}
  \bfalpha(\tau_j) = \bfalphah(\tau_j), \qquad j = 1,2,\ldots,k.
\end{equation}

Furthermore, since the linear systems \eqref{eq:MIMO} and \eqref{eq:MIMO-MK} have symmetric system matrices $(\bH,\bS)$ and $\bM\bK$, respectively, and the input and output matrices $\bfb$ and $\bfc$ are linearly dependent, the Galerkin projection becomes a Petrov--Galerkin projection\cite{Antoulas2005}. Hence, the original systems \eqref{eq:MIMO} and \eqref{eq:MIMO-MK} and its corresponding reduced order systems of dimension $k$ match $2k$ moments instead of only $k$ moments in the general case\cite{Antoulas2005}. In order words, we can obtain the same accuracy for the reduced order models with fewer interpolation frequencies than the general (non-linearly dependent) case.

\Algref{alg:mor} summarizes the construction of the reduced order model and how it is used to obtain an approximation of the absorption spectrum within an energy window of interest. Clearly, the higher the model order $k$, the more accurate the approximation. In the next section, we will show that even for a relatively small $k$, we can obtain a quite accurate approximation for $\sigma(\omega)$ in an interior spectral window that contains thousands of eigenvalues. 

\begin{algorithm2e}[hbtp]
\caption{Absorption spectrum via model order reduction}%
\label{alg:mor}%
\SetKwInOut{input}{Input}
\SetKwInOut{output}{Output}
\BlankLine
\input{Matrices $\bH,\bS,\bfd$, \\
       Interpolation frequencies $\tau_1,\tau_2\ldots,\tau_k$, \\
       Frequencies $\omega_1,\omega_2,\ldots,\omega_N$, and $\eta$.}
\BlankLine
\output{Absorption spectrum $\sigmah(\omega_1),\sigmah(\omega_2),\ldots,\sigmah(\omega_N)$.}
\BlankLine
\BlankLine
\For{$j=1,2,\ldots,k$}{
  \vspace{2pt}
  \nl Linear system solve $\bfx_j = \left( \bH - \tau_j\bS \right)\inv \bfd$.
}
\nl QR factorization $\bX = \bV\bR$.\\
\nl Construct $\bHh = \bV\T \bH \bV$, $\bSh = \bV\T \bS \bV$, and $\bfdh = \bV\T \bfd$. \\
\BlankLine
\For{$j=1,2,\ldots,N$}{
  \vspace{2pt}
  \nl Compute $\sigmah(\omega_j) = \omega \imag\left(\trace\left[ \bfdh\T \left( \bHh - (\omega_j + i\eta)\bSh \right)\inv \bfdh \right]\right)$.
}
\end{algorithm2e}

Although \algref{alg:mor} provides a general framework for constructing a reduced order model for estimating the absorption spectrum defined by $(\bH,\bS)$, it is more efficient to exploit the structure of $(\bH,\bS)$ and construct a reduced order model for \eqref{eq:MIMO-MK} instead. Such a reduced order model may be obtained by projecting \eqref{eq:MIMO-MK} onto a subspace defined by
\begin{equation}
  \bVt = \spn \Big[ \left( \bM\bK - \tau_1^2\bI \right)\inv \bfdt \quad
                    \left( \bM\bK - \tau_2^2\bI \right)\inv \bfdt \quad
                     \cdots \quad
                    \left( \bM\bK - \tau_k^2\bI \right)\inv \bfdt \Big],
  \label{eq:defV-MK}
\end{equation}
where $\tau_j$, $j = 1,2,\ldots,k$, are again the interpolation frequencies. Because the matrix $\bM\bK$ is self-adjoint with respect to the $\bK$-inner product, it is more convenient to carry out the projection using the $\bK$-inner product and projecting $\bM\bK$ onto a subspace spanned by a $\bK$-orthonormal basis, i.e., $\bVt\T \bK \bVt = \bI$ is satisfied. If we let
\begin{align}
  \bMKh &= \bVt\T \bK \bM \bK \bVt, \\
  \bfdh &= \bVt\T \bK \bfdt,
\end{align}
then the approximation to the absorption spectrum provided by the structure exploiting reduced order model can be expressed by
\begin{equation}
  \sigmah(\omega) \propto \omega \imag\left(\trace\left[ \bfdh\T \left( \bMKh - \omegat^2\bI \right)\inv \bfdh \right]\right).
  \label{eq:abs-spectrum-k-MK}
\end{equation}
By exploiting the block structure of $\bH$, we can prove that \eqref{eq:abs-spectrum-k} and \eqref{eq:abs-spectrum-k-MK} are equivalent. However, the latter is cheaper to construct, both in terms of the number of floating point operations and memory usage, since it only involves matrices of size $n\times n$ and vectors of size $n$. The structure exploiting model order reduction algorithm for approximating the absorption spectrum is outlined in \algref{alg:mor-MK}.

\begin{algorithm2e}[hbtp]
\caption{Absorption spectrum via structure exploiting model order reduction}%
\label{alg:mor-MK}%
\SetKwInOut{input}{Input}
\SetKwInOut{output}{Output}
\BlankLine
\input{Matrices $\bM,\bK,\bfdt$, \\
       Interpolation frequencies $\tau_1,\tau_2\ldots,\tau_k$, \\
       Frequencies $\omega_1,\omega_2,\ldots,\omega_N$, and $\eta$.}
\BlankLine
\output{Absorption spectrum $\sigmah(\omega_1),\sigmah(\omega_2),\ldots,\sigmah(\omega_N)$.}
\BlankLine
\BlankLine
\For{$j=1,2,\ldots,k$}{
  \vspace{2pt}
  \nl Linear system solve $\bfxt_j = \left( \bM\bK - \tau_j^2\bI \right)\inv \bfdt$.
}
\nl QR factorization $\bXt = \bVt\bRt$, with $\bVt\T \bK \bVt = \bI$.\\
\nl Construct $\bMKh = \bVt\T \bK\bM\bK \bVt$ and $\bfdh = \bVt\T \bK \bfdt$. \\
\BlankLine
\For{$j=1,2,\ldots,N$}{
  \vspace{2pt}
  \nl Compute $\sigmah(\omega_j) = \omega \imag\left(\trace\left[ \bfdh\T \left( \bMKh - (\omega_j + i\eta)^2\bI \right)\inv \bfdh \right]\right)$.
}
\end{algorithm2e}

Note that both \algref{alg:mor,alg:mor-MK} require a choice of the interpolation frequencies $\tau_j$. The number of these interpolation frequencies and their locations solely determine the quality of the absorption spectrum approximations. The simplest way to choose these interpolation frequencies is to partition the energy window of interest evenly by a uniform interpolation grid. However, because the absorption spectrum can be highly oscillatory in certain regions within the energy window, a very fine grid may be needed to resolve the high oscillation. As a result, the order of the reduced order model, which is proportional to the number of interpolation frequencies, can be exceedingly high.

\begin{figure}[hbtp]
\figname{adaptive}
\begin{tikzpicture}
\begin{axis}[%
 width=\textwidth,%
 height=0.5\textwidth,%
 axis x line=bottom,%
 axis y line=left,%
 xmin=0,%
 xmax=9,%
 ymin=0,%
 ymax=5,%
 xtick={0.5,...,8.5},%
 xticklabels={$\omega_\mathrm{min}$,,,,,,,,$\omega_\mathrm{max}$},%
 ytick={-1},%
 ylabel={error estimate},%
 grid=major,%
 label style={font=\small},%
 tick label style={font=\small},
]
\addplot[only marks,mark=square*]
  coordinates {
    (0.5,0)
    (2.5,0)
    (4.5,0)
    (6.5,0)
    (8.5,0)
  };
\addplot[only marks,mark=triangle*,mark size=3pt]
  coordinates {
    (1.5,0)
    (3.5,0)
    (5.5,0)
    (7.5,0)
  };
\addplot[only marks,mark=*,fill=white]
  coordinates {
    (1,0)
    (2,0)
    (3,0)
    (4,0)
    (5,0)
    (6,0)
    (7,0)
    (8,0)
  };
\addplot[only marks,mark=*] plot
  coordinates {
    (2,0)
    (3,0)
    (4,0)
    (7,0)
    (8,0)
  };
\legend{added in level 1,added in level 2,candidates for level 3,added in level 3}
\addplot[fill=white!95!black,draw=none] (1.51,0.01) -- (1.51,5) -- (2.49,5) -- (2.49,0.01);
\addplot[fill=white!95!black,draw=none] (2.51,0.01) -- (2.51,5) -- (3.49,5) -- (3.49,0.01);
\addplot[fill=white!95!black,draw=none] (3.51,0.01) -- (3.51,5) -- (4.49,5) -- (4.49,0.01);
\addplot[fill=white!95!black,draw=none] (6.51,0.01) -- (6.51,5) -- (7.49,5) -- (7.49,0.01);
\addplot[fill=white!95!black,draw=none] (7.51,0.01) -- (7.51,5) -- (8.49,5) -- (8.49,0.01);
\addplot[dotted,black] (2,0) -- (2,5);
\addplot[dotted,black] (3,0) -- (3,5);
\addplot[dotted,black] (4,0) -- (4,5);
\addplot[dotted,black] (7,0) -- (7,5);
\addplot[dotted,black] (8,0) -- (8,5);
\addplot[no marks,dashed,black] (0,3) -- (9,3);
\addplot[no marks] (0.5,3) node[above] {\scriptsize tolerance};
\addplot[blue,thick] expression[domain=0.5:8.5,samples=300] {2.5*sin(deg(x/2.25-0.5))*sin(2*deg(x/2.25-0.25)) + 2.5 + x/10};
\end{axis}
\end{tikzpicture}
\caption{Adaptive refinement strategy for selecting the interpolation frequencies.}
\label{fig:adaptive-strategy}
\end{figure}
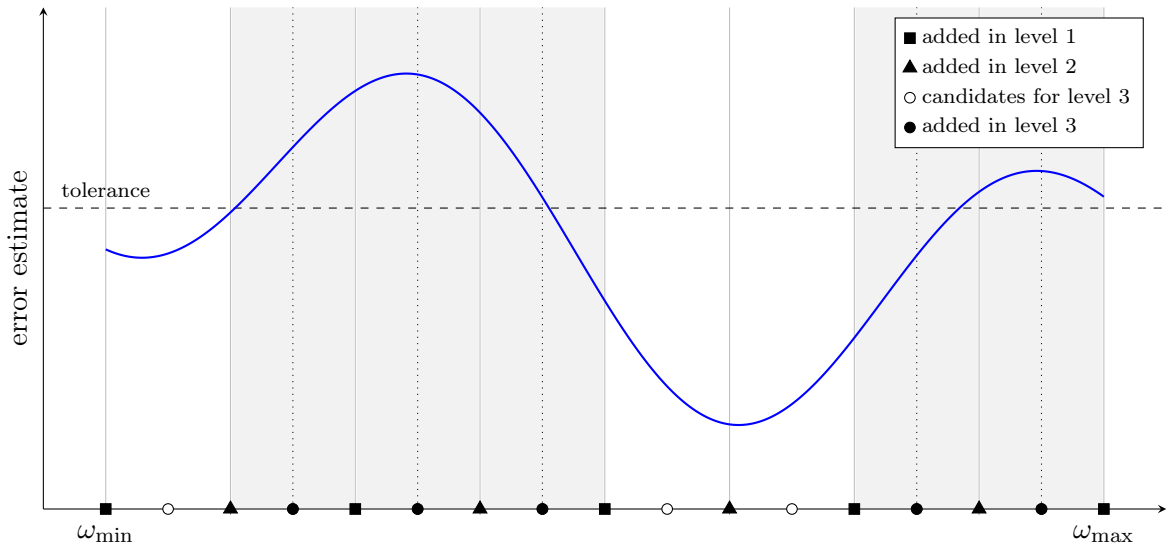

A more effective strategy for choosing the interpolation frequencies is to choose these frequencies in an adaptive fashion. We now propose a refinement strategy, which is graphically illustrated in \figref{fig:adaptive-strategy}. To start this procedure, we choose in the first level a coarse, uniform grid of interpolation frequencies (marked by {\scriptsize$\blacksquare$}) to construct the level-1 reduced order model. The set of interpolation frequencies is refined by adding the midpoints (marked by {\footnotesize$\blacktriangle$}) between two adjacent level-1 interpolation frequencies. This enlarged set forms the second level of interpolation frequencies, yielding a more accurate level-2 reduced order model. Next, we choose the midpoints between two adjacent level-2 interpolation frequencies as candidates (marked by $\circ$) to enlarge the set in the third level. We also estimate the approximation error by computing the relative difference between the level-1 and level-2 reduced order models for the entire energy window. If the error estimate at an interval between two adjacent level-2 interpolation frequencies is above a prescribed error tolerance, the midpoint (marked by $\bullet$) is added to the existing set of interpolation frequencies. The enlarge set results in an even more accurate level-3 reduced order model. This refinement process continues until the error estimate at the entire energy window is below the threshold or when the refined model order exceeds an prescribed upper bound.

\newcommand\plotspectrum[4][0.5]{%
\begin{tikzpicture}
\begin{axis}[%
 width=#1\textwidth,%
 xlabel={Energy (eV)},%
 ylabel={Absorption Spectrum (Arb.)},%
 xmin=540,xmax=600,%
 ymin=0,ymax=1,%
]
\addplot[thick,mark=,blue] table[x index=0,y index=#3]{\datfile{#2}};
\legend{{MOR (#4)}};
\end{axis}
\end{tikzpicture}%
}
\newcommand\plotspectrumos[5][0.5]{%
\begin{tikzpicture}
\begin{axis}[%
 width=#1\textwidth,%
 xlabel={Energy (eV)},%
 ylabel={Absorption Spectrum (Arb.)},%
 xmin=540,xmax=600,%
 ymin=0,ymax=1,%
 legend style={font=\tiny,fill=none,row sep=-2pt},%
]
\addplot[mark=,densely dotted] table[x index=0,y index=1]{\datfile{water_cluster_#5_refsol}};
\addplot[ycomb,gray] table[x index=0,y index=1]{\datfile{water_cluster_#5_oscstr}};
\addplot[thick,mark=,blue] table[x index=0,y index=#3]{\datfile{#2}};
\addplot[only marks,mark=+,red] table[x index=0,y index=1]{\datfile{water_cluster_#5_shifts}};
\legend{Eigensystem,,{MOR (#4)},Interpol.~freq.};
\end{axis}
\end{tikzpicture}%
}
\newcommand\plotfixed[2]{\plotspectrumos{real_vs_complex_shifts_fixed}{#1}{#2}{5H2O}}

\section{Computational results}
\label{sec:results}

The proposed automatic MOR algorithm has been implemented in the Chronus
Quantum software package\cite{chronusq_beta} and in
MATLAB\footnote[4]{\url{https://bitbucket.org/roelvb/mor4absspectrum}}.
The following numerical experiments were performed using a single
Sandy--Bridge Intel Xeon compute node (E5-2650 v2 @ 2.60 GHz) with 16
cores and 512 GB DDR3 RAM. All of the water cluster test cases were performed
using the 6-31G(d) basis set without the use of molecular symmetry and
were chosen for their dense spectral character in the X-Ray spectral domain.
All of the geometries for the water clusters used in this work may be
found in the supplemental information.

The implementation of the MOR utilizes a synchronized approach to the
Generalized Minimum Residual (GMRES)\cite{Walker88_152} algorithm for the
solution of the linear systems. In this approach \cite{shak2016}, each linear
system is solved individually via the standard GMRES algorithm but its
matrix-vector products (GEMVs), which constitutes the dominant cost, are synchronized
and performed in batches. Hence, the GEMVs become matrix-matrix
products (GEMMs) and allow for optimal efficiency and cache utilization through
the use of Level 3 BLAS operations. In all experiments we used a block size of
12, coming from combining the 3 dipole vectors at 4 interpolation frequencies.

Several numerical experiments were performed to demonstrate the performance and accuracy of the proposed MOR algorithms. Since the interpolation points are merely used to construct a reduced order model, it is conceivable that we may choose them to be real numbers instead of complex numbers that contain a small imaginary damping factor.  The advantage of choosing real interpolation points is that all linear systems can be solved in real arithmetic. However, as we will see below, this approach may not lead to any performance gain and can even lead to a performance degradation.

We also examined how the order of the reduced order model changes as the damping factor $\eta$ changes and as the size of the molecular system increases as well as the overall computational scaling of the proposed method using the aforementioned water clusters. Numerical comparisons are made to the Lorentzian broadened poles of the propagator using the oscillator strengths \cite{Ball64_844,Harris69_3947,McKoy75_1168}. The eigenvalues and oscillator strengths were computed via BSEPACK\cite{bsepack,SJYDL2016} on a Cray XC40 with Haswell Intel Xeon compute nodes (E5-2698 v3 @2.3 GHz, 2x16 cores, 128 GB DDR4 RAM). The broadening factor was set equal to $\eta$ for comparison with the approximate MOR experiments.

\subsection{Real versus complex interpolation frequencies}
\label{sec:results-points}

We start with a cluster of 5 water molecules and are interested in computing
the absorption spectrum in the energy window $[540\,\eV,600\,\eV]$. The
dimension of the matrix $\bH$ \eqref{eq:defH} was $2n = 6$,500 and $\bH$ had
394 eigenvalues in the energy window. The damping factor was $\eta = 1\,\eV$
and the tolerance for solving the linear systems was set to $10^{-6}$. The
damping factor was chosen to roughly mimic the effects of the core-hole
lifetime of the $K$-edge transitions in oxygen and vibrational
broadening\cite{Stohr_book}. It is important to note that the broadening due to
the damping parameter in these simulations is purely phenomenological, as no
vibronic effects are being explicitly treated.

\begin{figure}[hbtp]
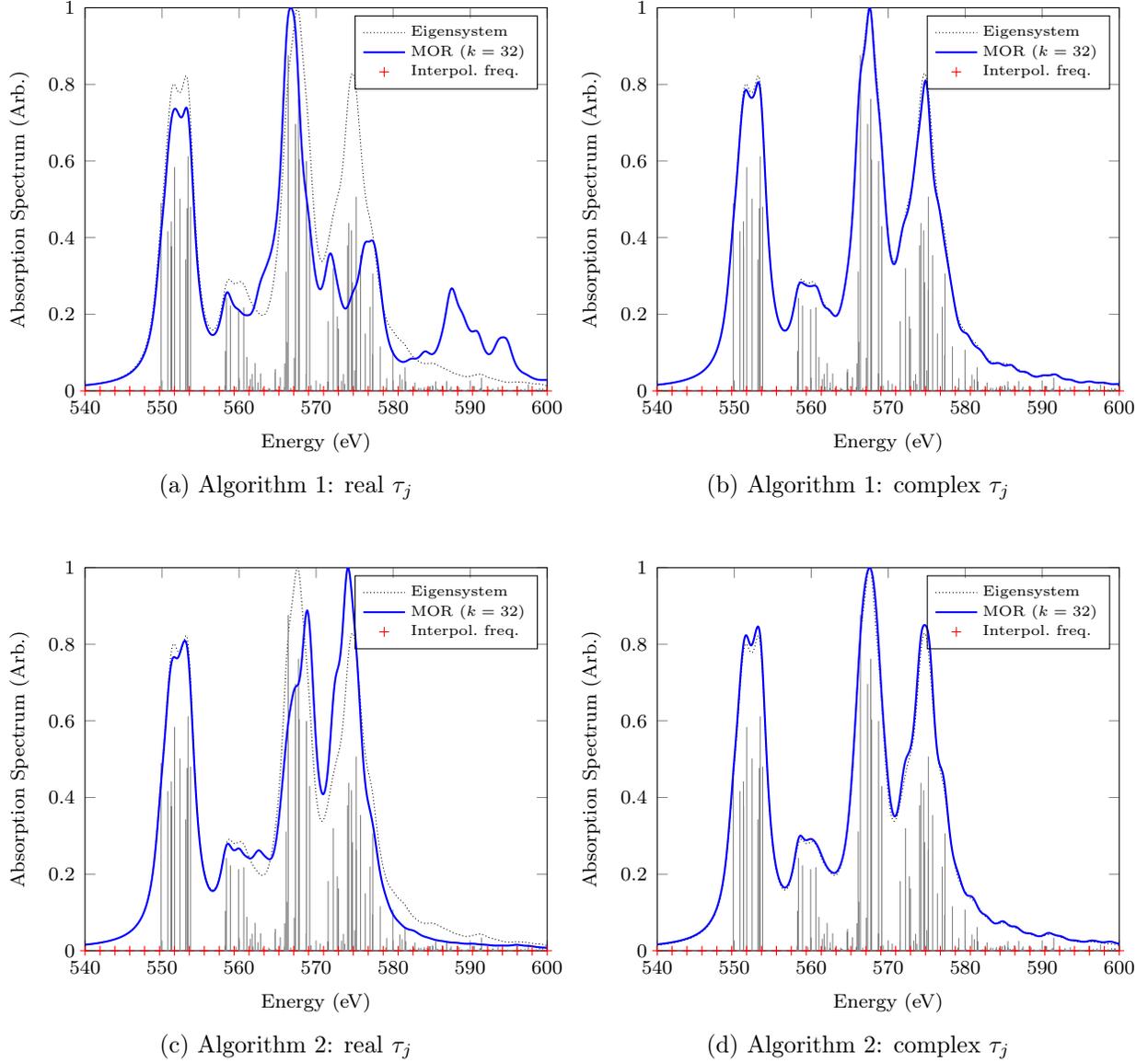

\fignames{fixed-}
\subfloat[\Algref{alg:mor}: real $\tau_j$]      {\plotfixed{2}{$k = 32$}}%
\subfloat[\Algref{alg:mor}: complex $\tau_j$]   {\plotfixed{3}{$k = 32$}}\\[10pt]
\subfloat[\Algref{alg:mor-MK}: real $\tau_j$]   {\plotfixed{4}{$k = 32$}}%
\subfloat[\Algref{alg:mor-MK}: complex $\tau_j$]{\plotfixed{5}{$k = 32$}}\\[10pt]
\caption{Numerical experiments for the evaluation of the XAS spectrum of 5 H$_2$O
clusters by the proposed MOR algorithms using a fixed model order ($k = 32$).
The MOR results are compared to the Lorentzian broadened poles of the
propagator, labelled Eigensystem. A damping parameter of $1\,\eV$ was chosen both
for the MOR calculations and the broadening factor of the Lorentzians for the
reference. It can be seen that the use of complex interpolation frequencies
for the construction of the model basis is important in spectrally dense
regions.}
\label{fig:fixed}
\end{figure}

In the first experiment, we used a fixed order $k = 32$ for the reduced order models and only changed the interpolation frequencies $\tau_j$, $j = 1,2,\ldots,k$. We computed the absorption spectrum by \algref{alg:mor,alg:mor-MK} for both real $\tau_j = \omega_j$ and complex $\tau_j = \omega_j + i\eta$, where $\omega_j$ were uniformly selected in the energy window. The corresponding results are presented in \figref{fig:fixed} and in the top part of \tabref{tab:real-vs-complex}. Note that by using complex interpolation frequencies $\tau_j$, we obtained  good approximations to the absorption spectrum from both \algref{alg:mor,alg:mor-MK} even with such a small model size. On the other hand, the use of real $\tau_j$ resulted in poor approximations for both algorithms. This is due to the fact that the (real) interpolation frequencies are often very close to the (real) eigenvalues of $(\bH,\bS)$ or $\bM\bK$, resulting in ill-conditioned linear systems to be solved. However, this can be avoided with complex interpolation frequencies.

\begin{table}[!b]
\caption{The effect of using real and complex interpolation frequencies $\tau_j$ on the MOR evaluation of XAS spectra for 5 H$_2$O clusters. Computational expense for \algref{alg:mor,alg:mor-MK}. Here $k$ is the reduced order, GEMMs is the total number of matrix-matrix products, and the total wall-clock time is given in seconds.%
\label{tab:real-vs-complex}}
\vspace{-0.5em}
\begin{center} \small
\begin{tabularx}{0.7\textwidth}{l|CrC|CrC|CrC}
\toprule
\multicolumn{1}{c|}{Algorithm} &
 \multicolumn{3}{c|}{$k$} &
 \multicolumn{3}{c|}{GEMMs} &
 \multicolumn{3}{c}{Wall (s)} \\
\midrule
\Algref{alg:mor}: real $\tau_j$       &&  32 &&& 1,052 &&& 19.76 & \\
\Algref{alg:mor}: complex $\tau_j$    &&  32 &&& 776 &&& 40.97 & \\
\Algref{alg:mor-MK}: real $\tau_j$    &&  32 &&& 985 &&& 9.78 & \\
\Algref{alg:mor-MK}: complex $\tau_j$ &&  32 &&& 646 &&& 17.5 & \\
\midrule
\Algref{alg:mor}: real $\tau_j$       && 218 &&& 7,440 &&& 137.01 & \\
\Algref{alg:mor}: complex $\tau_j$    &&  87 &&& 2,285 &&& 115.50 & \\
\Algref{alg:mor-MK}: real $\tau_j$    && 211 &&& 6,541 &&& 65.31 & \\
\Algref{alg:mor-MK}: complex $\tau_j$ &&  87 &&& 2,026 &&& 52.70 & \\
\midrule
\multicolumn{4}{l|}{Conventional CPP (1,000 points)}          &&   18,126   &&&   538.90   & \\
\bottomrule
\end{tabularx}
\vspace{-1em}
\end{center}
\end{table}

Next, we repeated the previous experiment but chose the interpolation
frequencies via the adaptive refinement strategy introduced in
\secref{sec:mor}. As the error estimates, we used the difference of
the normalized absorption spectrum between two consecutive refinement
levels. The tolerance was set to $0.01$, which corresponds to a 1
percent change in the overall absorption spectrum on the window
$[540\,\eV,600\,\eV]$. This resulted in reduced order models of different
orders $k$, reported in the middle part of
\tabref{tab:real-vs-complex}. We observe that in terms of the order
$k$, the use of complex interpolation frequencies has a significant advantage
over the use of real frequencies. Further, we also observe that the adaptive refinement strategy
for \algref{alg:mor,alg:mor-MK} resulted in very similar orders $k$ when the
same type of interpolation frequencies are used.

The corresponding computational expense for the previous two experiments is reported in \tabref{tab:real-vs-complex} using various metrics. We observe that for both fixed and adaptive model orders, the computational cost required for \algref{alg:mor-MK} was significantly lower than that of \algref{alg:mor}. This is expected as both methods are mathematically equivalent and the former only deals with linear systems of half the dimension of the latter. Furthermore, although \emph{real} interpolation frequencies allow us to solve only \emph{real} linear systems, we observe that in case of adaptively chosen model orders, the drastic decrease in model order required for complex interpolation frequencies over real frequencies offsets this advantage. Finally, we note at the bottom of \tabref{tab:real-vs-complex} that the use of \algref{alg:mor-MK} with complex interpolation frequencies reduces the computational expense by a factor of almost 10 compared to conventional complex polarization propagator calculations on a fine grid.

\subsection{Computational scaling}
\label{sec:results-scaling}

\begin{figure}[hbtp]
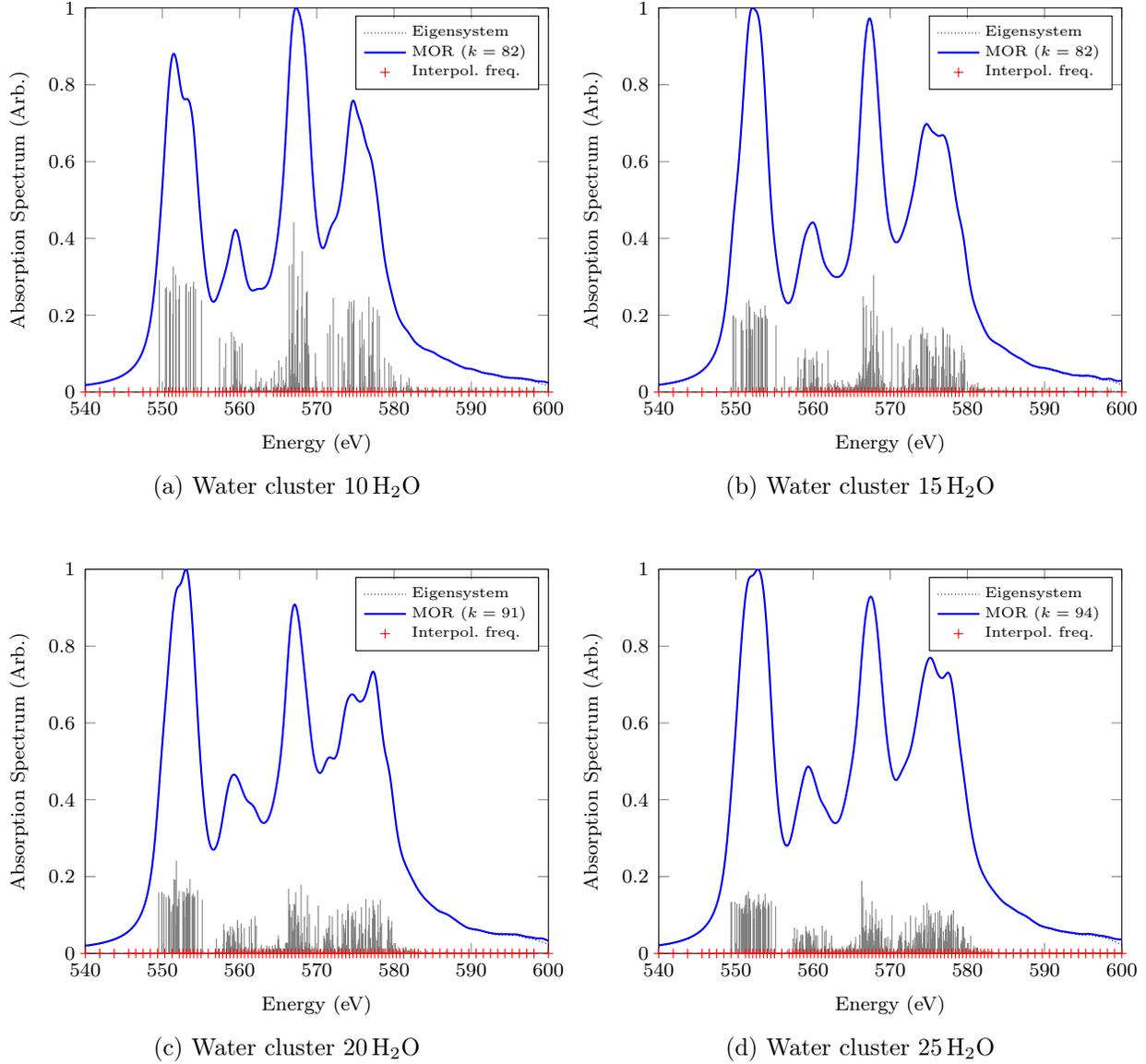

\centering
\figname{water_cluster_10H2O}%
\subfloat[Water cluster 10\,H$_2$O]{\plotspectrumos{water_cluster_10H2O_cq}{1}{$k = 82$}{10H2O}}
\figname{water_cluster_15H2O}%
\subfloat[Water cluster 15\,H$_2$O]{\plotspectrumos{water_cluster_15H2O_cq}{1}{$k = 82$}{15H2O}}\\[10pt]
\figname{water_cluster_20H2O}%
\subfloat[Water cluster 20\,H$_2$O]{\plotspectrumos{water_cluster_20H2O_cq}{1}{$k = 91$}{20H2O}}%
\figname{water_cluster_25H2O}%
\subfloat[Water cluster 25\,H$_2$O]{\plotspectrumos{water_cluster_25H2O_cq}{1}{$k = 94$}{25H2O}}\\[10pt]
\caption{Numerical experiments for the evaluation of the XAS spectrum of variably sized
H$_2$O clusters via \algref{alg:mor-MK} with adaptively chosen complex interpolation
frequencies. The MOR results are compared to the Lorentzian broadened poles of the
propagator, labelled Eigensystem. A damping parameter of $1\,\eV$ was chosen both
for the MOR calculations and the broadening factor of the Lorentzians for the
reference.}
\label{fig:water}
\end{figure}

We now consider water clusters consisting of 5, 10, 15, 20, and 25 water molecules. The corresponding matrix dimensions are shown in \tabref{tab:waters}. The energy window $[540\,\eV,600\,\eV]$ and damping factor $\eta = 1\,\eV$ were the same as for the previous experiments. We computed the absorption spectrum via \algref{alg:mor-MK} with complex interpolation frequencies chosen adaptively. The obtained absorption spectra are shown in \figref{fig:water}.

\begin{table}[hbtp]
\caption{Numerical experiments for the evaluation of the XAS spectrum of variably sized
H$_2$O clusters via \algref{alg:mor-MK} with adaptively chosen complex interpolation
frequencies. Here, $\bM\bK$ is of dimension $n$
with $\#\lambda$ eigenvalues lying within the energy window
$[540\,\eV,600\,\eV]$. The comparisons are made for GMRES convergence
tolerances of $10^{-4}$, $10^{-5}$, and $10^{-6}$, with $k$ as the reduced model order,
GEMMs as the total number of matrix-matrix products, and the total wall-clock time is given in
seconds.%
\label{tab:waters}}
\vspace{-0.5em}
\begin{center} \small
\begin{tabularx}{\textwidth}{rrr|rRr|rRr|rRr}
\toprule
\multicolumn{3}{c|}{Waters} &
 \multicolumn{3}{c|}{GMRES tol = $10^{-4}$} &
 \multicolumn{3}{c|}{GMRES tol = $10^{-5}$} &
 \multicolumn{3}{c}{GMRES tol = $10^{-6}$} \\[2pt]
\multicolumn{1}{c}{\#} &
 \multicolumn{1}{c}{$n$} &
 \multicolumn{1}{c|}{\#$\lambda$} &
 \multicolumn{3}{l|}{\ \,$k$ \hfill GEMMs \hfill Wall (s)} &
 \multicolumn{3}{l|}{\ \,$k$ \hfill GEMMs \hfill Wall (s)} &
 \multicolumn{3}{l}{\ \,$k$ \hfill GEMMs \hfill Wall (s)} \\
\midrule
 5 &  3,250 &   394 & \ 76 &   968 &     27.2 & \ 87 & 1,654 &     43.4 & \ 87 & 2,025 &     52.7 \\
10 & 13,000 & 1,456 &   99 & 1,749 &    636.2 &   83 & 2,404 &    867.1 &   82 & 3,235 &  1,157.0 \\
15 & 29,250 & 3,183 &   99 & 2,221 &  4,141.8 &   82 & 2,946 &  5,511.9 &   82 & 4,018 &  7,534.4 \\
20 & 52,000 & 5,524 &  123 & 2,742 & 14,665.8 &   89 & 3,317 & 17,807.0 &   91 & 4,594 & 25,656.5 \\
25 & 81,250 & 8,530 &  123 & 2,610 & 34,128.8 &   95 & 3,694 & 47,697.1 &   94 & 5,020 & 65,284.1 \\
\bottomrule
\end{tabularx}
\vspace{-1em}
\end{center}
\end{table}

The MOR results are given in \tabref{tab:waters}, where we present the orders $k$ of the reduced order models, the total number of GEMMs, and the total wall-clock time for different GMRES convergence tolerances. Firstly, we observe that the order $k$ of the reduced order models increases sub-linearly with the number of waters, whereas the number of eigenvalues inside the energy window, \#$\lambda$, grows linearly with respect to the problem dimension. Secondly, the order $k$ decreases for increasing GMRES convergence tolerances. This is due to the fact that if we solve the linear systems less accurately, we match the moments less accurately and hence we need more interpolation points (a higher value of $k$) for the same accuracy of the reduced order model and the corresponding absorption spectra. Moreover, the order $k$ seems to stagnate around GMRES tolerance $10^{-5}$ and there were no visual differences any more between the obtained absorption spectra for GMRES tolerances $10^{-5}$ and $10^{-6}$.

\begin{figure}[hbtp]
\centering
\subfloat[Wall time]{%
\figname{water_clusters_walltime}%
\begin{tikzpicture}
\begin{loglogaxis}[%
 width=0.49\textwidth,%
 xlabel={$n$},%
 ylabel={Wall (s)},%
 xmin=1e3,xmax=1e5,%
 ymin=1e1,ymax=1e5,%
 legend pos=north west,%
]
\addplot[thick,mark=*,blue]         table[x index=1,y index=4]{\datfile{water_clusters}};
\addplot[thick,mark=square*,red]    table[x index=1,y index=7]{\datfile{water_clusters}};
\addplot[thick,mark=triangle*,cyan] table[x index=1,y index=10]{\datfile{water_clusters}};
\addplot[no marks,gray,densely dotted] plot coordinates { (1e3,1e1) (1e5,1e3) };
\addplot[no marks,gray,densely dashed] plot coordinates { (1e3,1e1) (1e5,1e5) };
\addplot[no marks,gray] plot coordinates { (1e3,1e1) (1e5,1e7) };
\legend{$10^{-4}$,$10^{-5}$,$10^{-6}$,$\cO(n)$,$\cO(n^2)$,$\cO(n^3)$};
\end{loglogaxis}
\end{tikzpicture}%
}\hfill%
\subfloat[Total number of GEMMs]{%
\figname{water_clusters_gemms}%
\begin{tikzpicture}
\begin{loglogaxis}[%
 width=0.49\textwidth,%
 xlabel={$n$},%
 ylabel={GEMMs},%
 xmin=1e3,xmax=1e5,%
 ymin=1e2,ymax=1e6,%
 legend pos=north west,%
]
\addplot[thick,mark=*,blue]         table[x index=1,y index=3]{\datfile{water_clusters}};
\addplot[thick,mark=square*,red]    table[x index=1,y index=6]{\datfile{water_clusters}};
\addplot[thick,mark=triangle*,cyan] table[x index=1,y index=9]{\datfile{water_clusters}};
\addplot[no marks,gray,densely dotted] plot coordinates { (1e3,1e3) (1e5,1e5) };
\addplot[no marks,gray,densely dashdotted] table[x index=0,y index=1]{\datfile{O_log10}};
\legend{$10^{-4}$,$10^{-5}$,$10^{-6}$,$\cO(n)$,$\cO(\log_{10}(n))$};
\end{loglogaxis}
\end{tikzpicture}%
}%
\caption{Cluster of H$_2$O molecules: MOR results for the absorption spectra computed via \algref{alg:mor-MK} with adaptively chosen complex interpolation frequencies. The comparisons are made for GMRES convergence tolerances of $10^{-4}$, $10^{-5}$, and $10^{-6}$.}
\label{fig:scaling}
\end{figure}
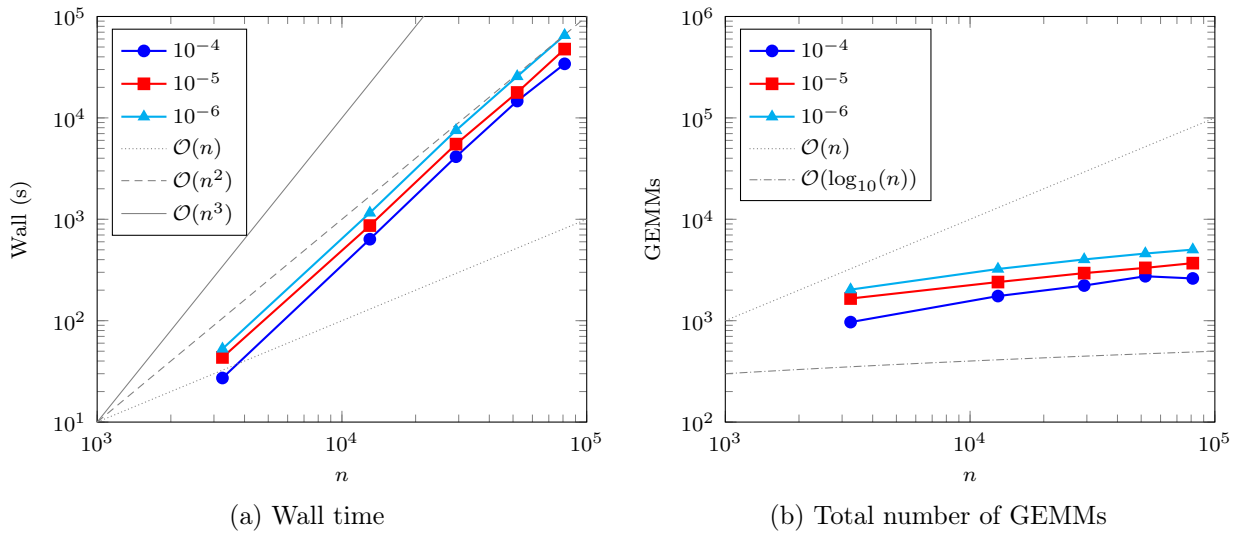

The total wall-clock time and number of GEMMs are also shown in \figref{fig:scaling}.
The left figure illustrates that the wall-clock time scales quadratically with
respect to the problem dimension, compared to a cubic scaling for a full
diagonalization. Moreover, the right figure shows that the number of GEMMs only
scales logarithmically, compared to an expected linear scaling for iterative
eigensolvers since the number of eigenvalues inside the energy window grows
linearly.
It is worth noting that the vector space dimension of the linear problem
also scales quadratically with system size.

\subsection{Effect of damping factor}
\label{sec:broad}

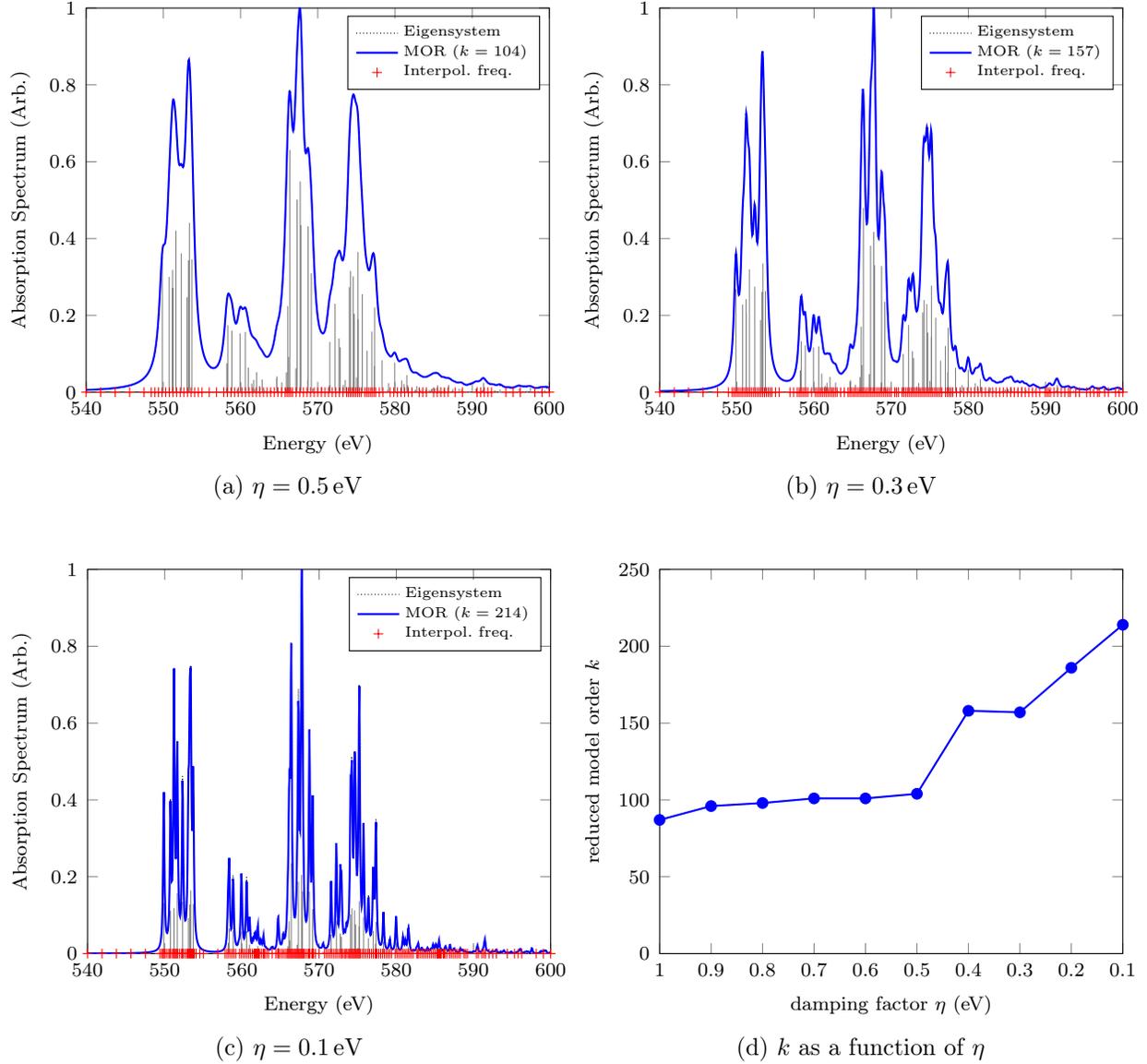
\begin{figure}[hbtp]
\centering
\figname{water_cluster_eta_0.5}%
\subfloat[$\eta = 0.5\,\eV$]{\plotspectrumos{water_cluster_eta_0.5}{1}{$k = 104$}{eta_0.5}}
\figname{water_cluster_eta_0.3}%
\subfloat[$\eta = 0.3\,\eV$]{\plotspectrumos{water_cluster_eta_0.3}{1}{$k = 157$}{eta_0.3}}\\[10pt]
\figname{water_cluster_eta_0.1}%
\subfloat[$\eta = 0.1\,\eV$]{\plotspectrumos{water_cluster_eta_0.1}{1}{$k = 214$}{eta_0.1}}
\figname{water_clusters_damping}%
\subfloat[$k$ as a function of $\eta$]{\begin{tikzpicture}
\begin{axis}[%
 width=0.5\textwidth,%
 xlabel={damping factor $\eta$ (eV)},%
 ylabel={reduced model order $k$},%
 xmin=0.1,xmax=1,%
 xtick={0.1,0.2,...,1},%
 ymin=0,ymax=250,%
 x dir=reverse,%
]
\addplot[thick,mark=*,blue] table[x index=0,y index=1]{\datfile{water_cluster_damping}};
\end{axis}
\end{tikzpicture}}\\[10pt]
\caption{Numerical experiments for the evaluation of the XAS spectrum of 5 H$_2$O clusters by \algref{alg:mor-MK} using different damping factors $\eta$. (a)--(c) The MOR results are compared to the Lorentzian broadened poles of the propagator, labelled Eigensystem. (d) Effect of the damping factor $\eta$ on the reduced model order $k$.}
\label{fig:damping}
\end{figure}

We examine the effect of the damping factor on the overall effectiveness of the proposed MOR algorithm in the low damping limit. We revisit the case of water clusters containing 5 water molecules from the previous subsections over the same energy widow. Specifically, we examine the effect on the damping parameter $\eta\in[0.1,1]\,\eV$ on the model order required to achieve a convergence of 1 percent in the absorption spectrum. The MOR results were obtains via \algref{alg:mor-MK} using adaptively chosen complex interpolation frequencies. The resulting spectra are presented in \figref{fig:damping}(a)--(c).

The effect of the damping factor on the automatically selected model order is illustrated in \figref{fig:damping}(d). In this figure, we observe that by decreasing the damping factor the reduced model order $k$ first remains almost constant until $0.5\,\eV$ and then slightly starts to increase for smaller values of $\eta$. Even in the low damping limit ($0.1\,\eV$), when the obtained absorption spectrum is exceptionally complicated and oscillatory relative to the previous experiments ($1\,\eV$), the required model order is still well within the realm of practicality for routine calculations. Thus the proposed MOR algorithm may be used as a general procedure which requires no assumption of (the smoothness of) the underlying absorption spectrum.

\section{Conclusion}
\label{sec:conclusion}
In this work, we have presented a novel, adaptive algorithm for the \emph{ab
initio} prediction of the absorption spectrum based on model order reduction
techniques applied to the quantum propagator. While this approach is general to
any spectral domain, the power of the proposed method is in those spectral
domains which are dense and interior in the propagator's eigenspectrum. The
accuracy and efficiency of this method to predict the X-Ray absorption spectrum
have been demonstrated using a series of water clusters. Water clusters were
chosen as an especially challenging case study as the propagator is spectrally
dense in the spectral neighborhood of the water's oxygen $K$-Edge. The
numerical experiments have shown that complex interpolation frequencies should
be preferred over real ones and that in this case the order of the reduced
order models only slightly increases with the problem dimension, in contrast to
the rapid growth of the number of eigenvalues inside the energy window.
Moreover, the wall-clock time for the proposed model order reduction algorithm scales
only quadratically with respect to the dimension of the problem,
compared to cubic scaling for eigenvalue based algorithms.
Further, it was shown that, even in the limit of highly oscillatory and low
damping absorption spectra, the proposed algorithm remains practical and thus
may be treated as agnostic to the underlying nature of the spectrum.
While results were presented only
for the TD-HF method, the proposed adaptive MOR algorithm is general to any
choice reference, propagator, or perturbation. Further, although it is not
expressly considered in this work, this technique is well suited for
parallelism on a massive scale as each of the linear system solutions is
completely independent from the other, thus allowing for minimal communication.
With the proposed MOR algorithm, routine study of X-Ray absorption spectra for
medium-to-large sized systems is simplified.

\begin{acknowledgement}
This work was partly supported by the Scientific Discovery through Advanced
Computing (SciDAC) program (R. V.\,B., C. Y. and E. G. N.) and the IDREAM
Energy Frontier Research Center (D. B. W.-Y., J. M. K. and X. L.), funded by
U.S.~Department of Energy, Office of Science, Advanced Scientific Computing
Research and Basic Energy Sciences. R. V.\,B. is a Postdoctoral Fellow of the
Research Foundation - Flanders (12J2217N). The development of the Chronus
Quantum software is supported by the National Science Foundation (CHE-1565520
to X. L.). The authors thank the National Energy Research Scientific Computing
(NERSC) center for making computational resources available to them. 
Further, this work was facilitated though the use of advanced
computational, storage, and networking infrastructure provided by the Hyak
supercomputer system at the University of Washington.
The authors are also grateful to Dr.~Meiyue Shao for helpful discussions.
\end{acknowledgement}

\bibliography{Journal_Short_Name,MORRefs}

\providecommand{\latin}[1]{#1}
\providecommand*\mcitethebibliography{\thebibliography}
\csname @ifundefined\endcsname{endmcitethebibliography}
  {\let\endmcitethebibliography\endthebibliography}{}
\begin{mcitethebibliography}{67}
\providecommand*\natexlab[1]{#1}
\providecommand*\mciteSetBstSublistMode[1]{}
\providecommand*\mciteSetBstMaxWidthForm[2]{}
\providecommand*\mciteBstWouldAddEndPuncttrue
  {\def\EndOfBibitem{\unskip.}}
\providecommand*\mciteBstWouldAddEndPunctfalse
  {\let\EndOfBibitem\relax}
\providecommand*\mciteSetBstMidEndSepPunct[3]{}
\providecommand*\mciteSetBstSublistLabelBeginEnd[3]{}
\providecommand*\EndOfBibitem{}
\mciteSetBstSublistMode{f}
\mciteSetBstMaxWidthForm{subitem}{(\alph{mcitesubitemcount})}
\mciteSetBstSublistLabelBeginEnd
  {\mcitemaxwidthsubitemform\space}
  {\relax}
  {\relax}

\bibitem[St{\"o}hr(2013)]{Stohr13_book}
St{\"o}hr,~J. \emph{NEXAFS spectroscopy}; Springer Science \& Business Media,
  2013; Vol.~25\relax
\mciteBstWouldAddEndPuncttrue
\mciteSetBstMidEndSepPunct{\mcitedefaultmidpunct}
{\mcitedefaultendpunct}{\mcitedefaultseppunct}\relax
\EndOfBibitem
\bibitem[Yannoulis \latin{et~al.}(1987)Yannoulis, Dudde, Frank, and
  Koch]{Koch87_519}
Yannoulis,~P.; Dudde,~R.; Frank,~K.; Koch,~E. Orientation of Aromatic
  Hydrocarbons on Metal Surfaces as Determined by NEXAFS. \emph{Surf. Sci.}
  \textbf{1987}, \emph{189}, 519--528\relax
\mciteBstWouldAddEndPuncttrue
\mciteSetBstMidEndSepPunct{\mcitedefaultmidpunct}
{\mcitedefaultendpunct}{\mcitedefaultseppunct}\relax
\EndOfBibitem
\bibitem[Ayg\"{u}l \latin{et~al.}(2012)Ayg\"{u}l, Batchelor, Dettinger, Yilmaz,
  Allard, Scherf, Peisert, and Chass\'{e}]{Chasse12_4870}
Ayg\"{u}l,~U.; Batchelor,~D.; Dettinger,~U.; Yilmaz,~S.; Allard,~S.;
  Scherf,~U.; Peisert,~H.; Chass\'{e},~T. Molecular Orientation in Polymer
  Films for Organic Solar Cells Studied by NEXAFS. \emph{J. Phys. Chem. C}
  \textbf{2012}, \emph{116}, 4870--4874\relax
\mciteBstWouldAddEndPuncttrue
\mciteSetBstMidEndSepPunct{\mcitedefaultmidpunct}
{\mcitedefaultendpunct}{\mcitedefaultseppunct}\relax
\EndOfBibitem
\bibitem[Shadle \latin{et~al.}(1995)Shadle, Hedman, Hodgson, and
  Solomon]{Solomon95_2259}
Shadle,~S.~E.; Hedman,~B.; Hodgson,~K.~O.; Solomon,~E.~I. Ligand K-Edge X-Ray
  Absorption Spectroscopic Studies. Metal-Ligand Covalency in a Series of
  Transition Metal Tetrachlorides. \emph{J. Am. Chem. Soc.} \textbf{1995},
  \emph{117}, 2259--2272\relax
\mciteBstWouldAddEndPuncttrue
\mciteSetBstMidEndSepPunct{\mcitedefaultmidpunct}
{\mcitedefaultendpunct}{\mcitedefaultseppunct}\relax
\EndOfBibitem
\bibitem[DuBois \latin{et~al.}(2000)DuBois, Mukherjee, Stack, Hedman, Solomon,
  and Hodgson]{Hodgson00_5775}
DuBois,~J.~L.; Mukherjee,~P.; Stack,~T.; Hedman,~B.; Solomon,~E.~I.;
  Hodgson,~K.~O. A Systematic K-Edge X-Ray Absorption Spectroscopic Study of Cu
  (III) Sites. \emph{J. Am. Chem. Soc.} \textbf{2000}, \emph{122},
  5775--5787\relax
\mciteBstWouldAddEndPuncttrue
\mciteSetBstMidEndSepPunct{\mcitedefaultmidpunct}
{\mcitedefaultendpunct}{\mcitedefaultseppunct}\relax
\EndOfBibitem
\bibitem[Chen \latin{et~al.}(2001)Chen, J{\"a}ger, Jennings, Gosztola,
  Munkholm, and Hessler]{Hessler01_262}
Chen,~L.~X.; J{\"a}ger,~W.~J.; Jennings,~G.; Gosztola,~D.~J.; Munkholm,~A.;
  Hessler,~J.~P. Capturing a photoexcited molecular structure through
  time-domain X-ray absorption fine structure. \emph{Science} \textbf{2001},
  \emph{292}, 262--264\relax
\mciteBstWouldAddEndPuncttrue
\mciteSetBstMidEndSepPunct{\mcitedefaultmidpunct}
{\mcitedefaultendpunct}{\mcitedefaultseppunct}\relax
\EndOfBibitem
\bibitem[Chen \latin{et~al.}(2016)Chen, Shelby, Lestrange, Jackson, Haldrup,
  Mara, Stickrath, Zhu, Lemke, Chollet, Hoffman, and Li]{Li16_639}
Chen,~L.~X.; Shelby,~M.~L.; Lestrange,~P.~J.; Jackson,~N.~E.; Haldrup,~K.;
  Mara,~M.~W.; Stickrath,~A.~B.; Zhu,~D.; Lemke,~H.; Chollet,~M.;
  Hoffman,~B.~M.; Li,~X. {Imaging Ultrafast Excited State Pathways in
  Transition Metal Complexes by X-ray Transient Absorption and Scattering using
  X-ray Free Electron Laser Source}. \emph{Faraday Discuss.} \textbf{2016},
  \emph{194}, 639--658\relax
\mciteBstWouldAddEndPuncttrue
\mciteSetBstMidEndSepPunct{\mcitedefaultmidpunct}
{\mcitedefaultendpunct}{\mcitedefaultseppunct}\relax
\EndOfBibitem
\bibitem[Shelby \latin{et~al.}(2016)Shelby, Lestrange, Jackson, Haldrup, Mara,
  Stickrath, Zhu, Lemke, Chollet, Hoffman, Li, and Chen]{Li16_JA2}
Shelby,~M.~L.; Lestrange,~P.~J.; Jackson,~N.~E.; Haldrup,~K.; Mara,~M.~W.;
  Stickrath,~A.~B.; Zhu,~D.; Lemke,~H.; Chollet,~M.; Hoffman,~B.~M.; Li,~X.;
  Chen,~L.~X. {Ultrafast Excited State Relaxation of a Metalloporphyrin
  Revealed by Femtosecond X-ray Absorption Spectroscopy}. \emph{J. Am. Chem.
  Soc.} \textbf{2016}, \emph{138}, 8752--8764\relax
\mciteBstWouldAddEndPuncttrue
\mciteSetBstMidEndSepPunct{\mcitedefaultmidpunct}
{\mcitedefaultendpunct}{\mcitedefaultseppunct}\relax
\EndOfBibitem
\bibitem[Li \latin{et~al.}(2005)Li, Smith, Markevitch, Romanov, Levis, and
  Schlegel]{Li05_233}
Li,~X.; Smith,~S.~M.; Markevitch,~A.~N.; Romanov,~D.~A.; Levis,~R.~J.;
  Schlegel,~H.~B. A Time-dependent Hartree-Fock Approach for Studying the
  Electronic Optical Response of Molecules in Intense Fields. \emph{Phys. Chem.
  Chem. Phys.} \textbf{2005}, \emph{7}, 233--239\relax
\mciteBstWouldAddEndPuncttrue
\mciteSetBstMidEndSepPunct{\mcitedefaultmidpunct}
{\mcitedefaultendpunct}{\mcitedefaultseppunct}\relax
\EndOfBibitem
\bibitem[Li and Tully(2007)Li, and Tully]{Li07_199}
Li,~X.; Tully,~J.~C. Ab initio Time-resolved Density Functional Theory for
  Lifetimes of Excited Adsorbate States at Metal Surfaces. \emph{Chem. Phys.
  Lett.} \textbf{2007}, \emph{439}, 199\relax
\mciteBstWouldAddEndPuncttrue
\mciteSetBstMidEndSepPunct{\mcitedefaultmidpunct}
{\mcitedefaultendpunct}{\mcitedefaultseppunct}\relax
\EndOfBibitem
\bibitem[Liang \latin{et~al.}(2011)Liang, Chapman, and Li]{Li11_184102}
Liang,~W.; Chapman,~C.~T.; Li,~X. Efficient First-principles Electronic
  Dynamics. \emph{J. Chem. Phys.} \textbf{2011}, \emph{134}, 184102\relax
\mciteBstWouldAddEndPuncttrue
\mciteSetBstMidEndSepPunct{\mcitedefaultmidpunct}
{\mcitedefaultendpunct}{\mcitedefaultseppunct}\relax
\EndOfBibitem
\bibitem[Lopata \latin{et~al.}(2012)Lopata, Van~Kuiken, Khalil, and
  Govind]{Govind12_3284}
Lopata,~K.; Van~Kuiken,~B.~E.; Khalil,~M.; Govind,~N. Linear-response and
  real-time time-dependent density functional theory studies of core-level
  near-edge x-ray absorption. \emph{J. Chem. Theor. Comput.} \textbf{2012},
  \emph{8}, 3284--3292\relax
\mciteBstWouldAddEndPuncttrue
\mciteSetBstMidEndSepPunct{\mcitedefaultmidpunct}
{\mcitedefaultendpunct}{\mcitedefaultseppunct}\relax
\EndOfBibitem
\bibitem[Bruner \latin{et~al.}(2016)Bruner, LaMaster, and
  Lopata]{Lopata16_3741}
Bruner,~A.; LaMaster,~D.; Lopata,~K. Accelerated broadband spectra using
  transition dipole decomposition and Pad{\'e} approximants. \emph{J. Chem.
  Theor. Comput.} \textbf{2016}, \emph{12}, 3741--3750\relax
\mciteBstWouldAddEndPuncttrue
\mciteSetBstMidEndSepPunct{\mcitedefaultmidpunct}
{\mcitedefaultendpunct}{\mcitedefaultseppunct}\relax
\EndOfBibitem
\bibitem[Oddershede \latin{et~al.}(1984)Oddershede, J{\o}rgensen, and
  Yeager]{Yeager84_33}
Oddershede,~J.; J{\o}rgensen,~P.; Yeager,~D.~L. Polarization Propagator Methods
  in Atomic and Molecular Calculations. \emph{Comp. Phys. Rep.} \textbf{1984},
  \emph{2}, 33--92\relax
\mciteBstWouldAddEndPuncttrue
\mciteSetBstMidEndSepPunct{\mcitedefaultmidpunct}
{\mcitedefaultendpunct}{\mcitedefaultseppunct}\relax
\EndOfBibitem
\bibitem[Olsen \latin{et~al.}(1988)Olsen, Jensen, and J{\o}rgensen]{olje1988}
Olsen,~J.; Jensen,~H. J.~A.; J{\o}rgensen,~P. {Solution of the large matrix
  equations which occur in response theory}. \emph{J. Comput. Phys.}
  \textbf{1988}, \emph{74}, 265--282\relax
\mciteBstWouldAddEndPuncttrue
\mciteSetBstMidEndSepPunct{\mcitedefaultmidpunct}
{\mcitedefaultendpunct}{\mcitedefaultseppunct}\relax
\EndOfBibitem
\bibitem[Monkhorst(1977)]{Monkhorst77_421}
Monkhorst,~H.~J. Calculation of Properties with the Coupled-Cluster Method.
  \emph{Int. J. Quant. Chem.} \textbf{1977}, \emph{12}, 421--432\relax
\mciteBstWouldAddEndPuncttrue
\mciteSetBstMidEndSepPunct{\mcitedefaultmidpunct}
{\mcitedefaultendpunct}{\mcitedefaultseppunct}\relax
\EndOfBibitem
\bibitem[Koch and J{\o}rgensen(1990)Koch, and J{\o}rgensen]{Jorgensen90_3333}
Koch,~H.; J{\o}rgensen,~P. Coupled Cluster Response Functions. \emph{J. Chem.
  Phys.} \textbf{1990}, \emph{93}, 3333--3344\relax
\mciteBstWouldAddEndPuncttrue
\mciteSetBstMidEndSepPunct{\mcitedefaultmidpunct}
{\mcitedefaultendpunct}{\mcitedefaultseppunct}\relax
\EndOfBibitem
\bibitem[Stanton and Bartlett(1993)Stanton, and Bartlett]{Bartlett93_7029}
Stanton,~J.~F.; Bartlett,~R.~J. The Equation of Motion Coupled-Cluster Method.
  A Systematic Biorthogonal Approach to Molecular Excitation Energies,
  Transition Probabilities, and Excited State Properties. \emph{J. Chem. Phys.}
  \textbf{1993}, \emph{98}, 7029--7039\relax
\mciteBstWouldAddEndPuncttrue
\mciteSetBstMidEndSepPunct{\mcitedefaultmidpunct}
{\mcitedefaultendpunct}{\mcitedefaultseppunct}\relax
\EndOfBibitem
\bibitem[Comeau and Bartlett(1993)Comeau, and Bartlett]{Bartlett93_414}
Comeau,~D.~C.; Bartlett,~R.~J. The Equation-of-Motion Coupled-Cluster Method.
  Applications to Open-and Closed-Shell Reference States. \emph{Chem. Phys.
  Lett.} \textbf{1993}, \emph{207}, 414--423\relax
\mciteBstWouldAddEndPuncttrue
\mciteSetBstMidEndSepPunct{\mcitedefaultmidpunct}
{\mcitedefaultendpunct}{\mcitedefaultseppunct}\relax
\EndOfBibitem
\bibitem[Wenzel \latin{et~al.}(2014)Wenzel, Wormit, and Dreuw]{Dreuw14_4583}
Wenzel,~J.; Wormit,~M.; Dreuw,~A. {Calculating X-ray absorption spectra of
  open-shell molecules with the unrestricted algebraic-diagrammatic
  construction scheme for the polarization propagator}. \emph{J. Chem. Theor.
  Comput.} \textbf{2014}, \emph{10}, 4583\relax
\mciteBstWouldAddEndPuncttrue
\mciteSetBstMidEndSepPunct{\mcitedefaultmidpunct}
{\mcitedefaultendpunct}{\mcitedefaultseppunct}\relax
\EndOfBibitem
\bibitem[Wenzel \latin{et~al.}(2014)Wenzel, Wormit, and Dreuw]{Dreuw14_1900}
Wenzel,~J.; Wormit,~M.; Dreuw,~A. {Calculating core-level excitations and X-ray
  absorption spectra of large and medium sized closed-shell molecules with the
  algebraic-diagrammatic construction scheme for the polarization propagator}.
  \emph{J. Chem. Theor. Comput.} \textbf{2014}, \emph{35}, 1900\relax
\mciteBstWouldAddEndPuncttrue
\mciteSetBstMidEndSepPunct{\mcitedefaultmidpunct}
{\mcitedefaultendpunct}{\mcitedefaultseppunct}\relax
\EndOfBibitem
\bibitem[Christiansen \latin{et~al.}(1998)Christiansen, J{\o}rgensen, and
  H{\"a}ttig]{Hattig98_1}
Christiansen,~O.; J{\o}rgensen,~P.; H{\"a}ttig,~C. Response Functions from
  Fourier Component Variational Perturbation Theory Applied to a Time-Averaged
  Quasienergy. \emph{Int. J. Quant. Chem.} \textbf{1998}, \emph{68},
  1--52\relax
\mciteBstWouldAddEndPuncttrue
\mciteSetBstMidEndSepPunct{\mcitedefaultmidpunct}
{\mcitedefaultendpunct}{\mcitedefaultseppunct}\relax
\EndOfBibitem
\bibitem[Ring and Schuck(2004)Ring, and Schuck]{Ring_book}
Ring,~P.; Schuck,~P. \emph{The Nuclear Many-Body Problem}; Springer Science \&
  Business Media, 2004\relax
\mciteBstWouldAddEndPuncttrue
\mciteSetBstMidEndSepPunct{\mcitedefaultmidpunct}
{\mcitedefaultendpunct}{\mcitedefaultseppunct}\relax
\EndOfBibitem
\bibitem[J{\o}rgensen and Simmons(1981)J{\o}rgensen, and
  Simmons]{Jorgensen_book}
J{\o}rgensen,~P.; Simmons,~J. \emph{Second Quantization--Based Methods in
  Quantum Chemistry}; Academic Press Inc., 1981\relax
\mciteBstWouldAddEndPuncttrue
\mciteSetBstMidEndSepPunct{\mcitedefaultmidpunct}
{\mcitedefaultendpunct}{\mcitedefaultseppunct}\relax
\EndOfBibitem
\bibitem[Rowe(1968)]{Rowe68_153}
Rowe,~D. Equations-of-Motion Method and the Extended Shell Model. \emph{Rev.
  Mod. Phys.} \textbf{1968}, \emph{40}, 153\relax
\mciteBstWouldAddEndPuncttrue
\mciteSetBstMidEndSepPunct{\mcitedefaultmidpunct}
{\mcitedefaultendpunct}{\mcitedefaultseppunct}\relax
\EndOfBibitem
\bibitem[Casida(1995)]{Casida95_Book}
Casida,~M.~E. In \emph{Recent Advances in Density Functional Methods:(Part I)};
  Chong,~D.~P., Ed.; World Scientific; Singapore, 1995; Vol.~1; pp
  155--193\relax
\mciteBstWouldAddEndPuncttrue
\mciteSetBstMidEndSepPunct{\mcitedefaultmidpunct}
{\mcitedefaultendpunct}{\mcitedefaultseppunct}\relax
\EndOfBibitem
\bibitem[Dreuw and Head-Gordon(2005)Dreuw, and Head-Gordon]{HeadGordon05_4009}
Dreuw,~A.; Head-Gordon,~M. Single-Reference Ab Initio Methods for the
  Calculation of Excited States of Large Molecules. \emph{Chem. Rev.}
  \textbf{2005}, \emph{105}, 4009--4037\relax
\mciteBstWouldAddEndPuncttrue
\mciteSetBstMidEndSepPunct{\mcitedefaultmidpunct}
{\mcitedefaultendpunct}{\mcitedefaultseppunct}\relax
\EndOfBibitem
\bibitem[Helgaker \latin{et~al.}(2012)Helgaker, Coriani, J{\o}rgensen,
  Kristensen, Olsen, and Ruud]{Ruud12_543}
Helgaker,~T.; Coriani,~S.; J{\o}rgensen,~P.; Kristensen,~K.; Olsen,~J.;
  Ruud,~K. Recent Advances in Wave Function-Based Methods of Molecular-Property
  Calculations. \emph{Chem. Rev.} \textbf{2012}, \emph{112}, 543--631\relax
\mciteBstWouldAddEndPuncttrue
\mciteSetBstMidEndSepPunct{\mcitedefaultmidpunct}
{\mcitedefaultendpunct}{\mcitedefaultseppunct}\relax
\EndOfBibitem
\bibitem[Shavitt and Bartlett(2009)Shavitt, and Bartlett]{Bartlett09_Book}
Shavitt,~I.; Bartlett,~R.~J. \emph{Many-Body Methods in Chemistry and Physics:
  MBPT and Coupled-Cluster Theory}; Cambridge university press, 2009\relax
\mciteBstWouldAddEndPuncttrue
\mciteSetBstMidEndSepPunct{\mcitedefaultmidpunct}
{\mcitedefaultendpunct}{\mcitedefaultseppunct}\relax
\EndOfBibitem
\bibitem[Liang \latin{et~al.}(2011)Liang, Fischer, Frisch, and Li]{Li11_3540}
Liang,~W.; Fischer,~S.~A.; Frisch,~M.~J.; Li,~X. Energy-Specific Linear
  Response TDHF/TDDFT for Calculating High-Energy Excited States. \emph{J.
  Chem. Theor. Comput.} \textbf{2011}, \emph{7}, 3540--3547\relax
\mciteBstWouldAddEndPuncttrue
\mciteSetBstMidEndSepPunct{\mcitedefaultmidpunct}
{\mcitedefaultendpunct}{\mcitedefaultseppunct}\relax
\EndOfBibitem
\bibitem[Lestrange \latin{et~al.}(2015)Lestrange, Nguyen, and Li]{Li15_2994}
Lestrange,~P.~J.; Nguyen,~P.~D.; Li,~X. {Calibration of Energy-Specific TDDFT
  for Modeling K-edge XAS Spectra of Light Elements}. \emph{J. Chem. Theor.
  Comput.} \textbf{2015}, \emph{11}, 2994--2999\relax
\mciteBstWouldAddEndPuncttrue
\mciteSetBstMidEndSepPunct{\mcitedefaultmidpunct}
{\mcitedefaultendpunct}{\mcitedefaultseppunct}\relax
\EndOfBibitem
\bibitem[Peng \latin{et~al.}(2015)Peng, Lestrange, Goings, Caricato, and
  Li]{Li15_4146}
Peng,~B.; Lestrange,~P.~J.; Goings,~J.~J.; Caricato,~M.; Li,~X.
  {Energy-Specific Equation-of-Motion Coupled-Cluster Methods for High-Energy
  Excited States: Application to K-edge X-ray Absorption Spectroscopy}.
  \emph{J. Chem. Theor. Comput.} \textbf{2015}, \emph{11}, 4146--4153\relax
\mciteBstWouldAddEndPuncttrue
\mciteSetBstMidEndSepPunct{\mcitedefaultmidpunct}
{\mcitedefaultendpunct}{\mcitedefaultseppunct}\relax
\EndOfBibitem
\bibitem[Stener \latin{et~al.}(2003)Stener, Fronzoni, and
  De~Simone]{DeSimone03_115}
Stener,~M.; Fronzoni,~G.; De~Simone,~M. Time Dependent Density Functional
  Theory of Core Electrons Excitations. \emph{Chem. Phys. Lett.} \textbf{2003},
  \emph{373}, 115--123\relax
\mciteBstWouldAddEndPuncttrue
\mciteSetBstMidEndSepPunct{\mcitedefaultmidpunct}
{\mcitedefaultendpunct}{\mcitedefaultseppunct}\relax
\EndOfBibitem
\bibitem[Ray \latin{et~al.}(2007)Ray, DeBeer~George, Solomon, Wieghardt, and
  Neese]{Neese07_2783}
Ray,~K.; DeBeer~George,~S.; Solomon,~E.~I.; Wieghardt,~K.; Neese,~F.
  Description of the Ground-State Covalencies of the Bis (dithiolato)
  Transition-Metal Complexes from X-ray Absorption Spectroscopy and
  Time-Dependent Density-Functional Calculations. \emph{Chem. Eur. J.}
  \textbf{2007}, \emph{13}, 2783--2797\relax
\mciteBstWouldAddEndPuncttrue
\mciteSetBstMidEndSepPunct{\mcitedefaultmidpunct}
{\mcitedefaultendpunct}{\mcitedefaultseppunct}\relax
\EndOfBibitem
\bibitem[Besley and Asmuruf(2010)Besley, and Asmuruf]{Asmuruf10_12024}
Besley,~N.~A.; Asmuruf,~F.~A. Time-Dependent Density Functional Theory
  Calculations of the Spectroscopy of Core Electrons. \emph{Phys. Chem. Chem.
  Phys.} \textbf{2010}, \emph{12}, 12024--12039\relax
\mciteBstWouldAddEndPuncttrue
\mciteSetBstMidEndSepPunct{\mcitedefaultmidpunct}
{\mcitedefaultendpunct}{\mcitedefaultseppunct}\relax
\EndOfBibitem
\bibitem[Davidson(1975)]{Davidson75_87}
Davidson,~E.~R. {The Iterative Calculation of a Few of the Lowest Eigenvalues
  and Corresponding Eigenvectors of Large Real--Symmetric Matricies}. \emph{J.
  Comput. Phys.} \textbf{1975}, \emph{17}, 87\relax
\mciteBstWouldAddEndPuncttrue
\mciteSetBstMidEndSepPunct{\mcitedefaultmidpunct}
{\mcitedefaultendpunct}{\mcitedefaultseppunct}\relax
\EndOfBibitem
\bibitem[Morgan and Scott(1986)Morgan, and Scott]{Scott86_817}
Morgan,~R.~B.; Scott,~D.~S. {Generalizations of Davidson's Method for Computing
  Eigenvalues of Sparse Symmetric Matricies}. \emph{{SIAM J. Sci. Statist.
  Comput.}} \textbf{1986}, \emph{7}, 817--825\relax
\mciteBstWouldAddEndPuncttrue
\mciteSetBstMidEndSepPunct{\mcitedefaultmidpunct}
{\mcitedefaultendpunct}{\mcitedefaultseppunct}\relax
\EndOfBibitem
\bibitem[Morgan(1992)]{Morgan92_287}
Morgan,~R.~B. {Generalizations of Davidson's Method for Computing Eigenvalues
  of Large Non--Symmetric Matricies}. \emph{J. Comput. Phys.} \textbf{1992},
  \emph{101}, 287--291\relax
\mciteBstWouldAddEndPuncttrue
\mciteSetBstMidEndSepPunct{\mcitedefaultmidpunct}
{\mcitedefaultendpunct}{\mcitedefaultseppunct}\relax
\EndOfBibitem
\bibitem[Zuev \latin{et~al.}(2015)Zuev, Vecharynski, Yang, Orms, and
  Krylov]{zuev_etal2015}
Zuev,~D.; Vecharynski,~E.; Yang,~C.; Orms,~N.; Krylov,~A.~I. New Algorithms for
  Iterative Matrix-Free Eigensolvers in Quantum Chemistry. \emph{J. Comput.
  Chem.} \textbf{2015}, \emph{36}, 273--284\relax
\mciteBstWouldAddEndPuncttrue
\mciteSetBstMidEndSepPunct{\mcitedefaultmidpunct}
{\mcitedefaultendpunct}{\mcitedefaultseppunct}\relax
\EndOfBibitem
\bibitem[Coriani \latin{et~al.}(2012)Coriani, Fransson, Christiansen, and
  Norman]{Norman12_1616}
Coriani,~S.; Fransson,~T.; Christiansen,~O.; Norman,~P.
  Asymmetric-Lanczos-chain-driven implementation of electronic resonance
  convergent coupled-cluster linear response theory. \emph{J. Chem. Theor.
  Comput.} \textbf{2012}, \emph{8}, 1616--1628\relax
\mciteBstWouldAddEndPuncttrue
\mciteSetBstMidEndSepPunct{\mcitedefaultmidpunct}
{\mcitedefaultendpunct}{\mcitedefaultseppunct}\relax
\EndOfBibitem
\bibitem[Fransson \latin{et~al.}(2013)Fransson, Coriani, Christiansen, and
  Norman]{Norman13_124311}
Fransson,~T.; Coriani,~S.; Christiansen,~O.; Norman,~P. Carbon X-ray Absorption
  Spectra of Fluoroethenes and Acetone: A Study at the Coupled Cluster, Density
  Functional, and Static-exchange Levels of Theory. \emph{J. Chem. Phys.}
  \textbf{2013}, \emph{138}, 124311\relax
\mciteBstWouldAddEndPuncttrue
\mciteSetBstMidEndSepPunct{\mcitedefaultmidpunct}
{\mcitedefaultendpunct}{\mcitedefaultseppunct}\relax
\EndOfBibitem
\bibitem[Kauczor \latin{et~al.}(2013)Kauczor, Norman, Christiansen, and
  Coriani]{Coriani13_211102}
Kauczor,~J.; Norman,~P.; Christiansen,~O.; Coriani,~S. Communication: A
  Reduced-Space Algorithm for the Solution of the Complex Linear Response
  Equations used in Coupled Cluster Damped Response Theory. 2013\relax
\mciteBstWouldAddEndPuncttrue
\mciteSetBstMidEndSepPunct{\mcitedefaultmidpunct}
{\mcitedefaultendpunct}{\mcitedefaultseppunct}\relax
\EndOfBibitem
\bibitem[Marques \latin{et~al.}(2012)Marques, Maitra, Nogueira, Gross, and
  Rubio]{Rubio_Book}
Marques,~M.~A.; Maitra,~N.~T.; Nogueira,~F.~M.; Gross,~E.~K.; Rubio,~A.
  \emph{Fundamentals of Time-Dependent Density Functional Theory}; Springer
  Science \& Business Media, 2012; Vol. 837; Chapter 7\relax
\mciteBstWouldAddEndPuncttrue
\mciteSetBstMidEndSepPunct{\mcitedefaultmidpunct}
{\mcitedefaultendpunct}{\mcitedefaultseppunct}\relax
\EndOfBibitem
\bibitem[Norman \latin{et~al.}(2001)Norman, Bishop, J{\o}rgen Aa.~Jensen, and
  Oddershede]{Oddershede01_JCP}
Norman,~P.; Bishop,~D.~M.; J{\o}rgen Aa.~Jensen,~H.; Oddershede,~J.
  Near-Resonant Absorption in the Time-Dependent Self-Consistent Field and
  Multiconfigurational Self-Consistent Field Approximations. \emph{J. Chem.
  Phys.} \textbf{2001}, \emph{115}, 10323--10334\relax
\mciteBstWouldAddEndPuncttrue
\mciteSetBstMidEndSepPunct{\mcitedefaultmidpunct}
{\mcitedefaultendpunct}{\mcitedefaultseppunct}\relax
\EndOfBibitem
\bibitem[Fahleson \latin{et~al.}(2016)Fahleson, {\AA}gren, and
  Norman]{Norman16_1991}
Fahleson,~T.; {\AA}gren,~H.; Norman,~P. A Polarization Propagator for Nonlinear
  X-ray Spectroscopies. \emph{J. Phys. Chem. Lett.} \textbf{2016}, \emph{7},
  1991--1995\relax
\mciteBstWouldAddEndPuncttrue
\mciteSetBstMidEndSepPunct{\mcitedefaultmidpunct}
{\mcitedefaultendpunct}{\mcitedefaultseppunct}\relax
\EndOfBibitem
\bibitem[Coriani \latin{et~al.}(2012)Coriani, Christiansen, Fransson, and
  Norman]{Norman12_022507}
Coriani,~S.; Christiansen,~O.; Fransson,~T.; Norman,~P. Coupled-Cluster
  Response Theory for Near-Edge X-Ray-Absorption Fine Structure of Atoms and
  Molecules. \emph{Phys. Rev. A} \textbf{2012}, \emph{85}, 022507\relax
\mciteBstWouldAddEndPuncttrue
\mciteSetBstMidEndSepPunct{\mcitedefaultmidpunct}
{\mcitedefaultendpunct}{\mcitedefaultseppunct}\relax
\EndOfBibitem
\bibitem[Linares \latin{et~al.}(2010)Linares, Stafström, Rinkevicius,
  {\AA}gren, and Norman]{Norman10_5096}
Linares,~M.; Stafström,~S.; Rinkevicius,~Z.; {\AA}gren,~H.; Norman,~P.
  Complex Polarization Propagator Approach in the Restricted Open-Shell,
  Self-Consistent Field Approximation: The Near K-Edge X-ray Absorption Fine
  Structure Spectra of Allyl and Copper Phthalocyanine. \emph{J. Phys. Chem. B}
  \textbf{2010}, \emph{115}, 5096--5102\relax
\mciteBstWouldAddEndPuncttrue
\mciteSetBstMidEndSepPunct{\mcitedefaultmidpunct}
{\mcitedefaultendpunct}{\mcitedefaultseppunct}\relax
\EndOfBibitem
\bibitem[Ekstr{\"o}m \latin{et~al.}(2006)Ekstr{\"o}m, Norman, Carravetta, and
  {\AA}gren]{Agren06_143001}
Ekstr{\"o}m,~U.; Norman,~P.; Carravetta,~V.; {\AA}gren,~H. Polarization
  Propagator for X-ray Spectra. \emph{Phys. Rev. Lett.} \textbf{2006},
  \emph{97}, 143001\relax
\mciteBstWouldAddEndPuncttrue
\mciteSetBstMidEndSepPunct{\mcitedefaultmidpunct}
{\mcitedefaultendpunct}{\mcitedefaultseppunct}\relax
\EndOfBibitem
\bibitem[Fransson \latin{et~al.}(2016)Fransson, Burdakova, and
  Norman]{Norman16_13591}
Fransson,~T.; Burdakova,~D.; Norman,~P. K-and L-Edge X-Ray Absorption Spectrum
  Calculations of Closed-Shell Carbon, Silicon, Germanium, and Sulfur Compounds
  using Damped Four-Component Density Functional Response Theory. \emph{Phys.
  Chem. Chem. Phys.} \textbf{2016}, \emph{18}, 13591--13603\relax
\mciteBstWouldAddEndPuncttrue
\mciteSetBstMidEndSepPunct{\mcitedefaultmidpunct}
{\mcitedefaultendpunct}{\mcitedefaultseppunct}\relax
\EndOfBibitem
\bibitem[Norman \latin{et~al.}(2015)Norman, Parello, Polavarapu, and
  Linares]{Mathieu15_21866}
Norman,~P.; Parello,~J.; Polavarapu,~P.~L.; Linares,~M. Predicting Near-UV
  Electronic Circular Dichroism in Nucleosomal DNA by Means of DFT Response
  Theory. \emph{Phys. Chem. Chem. Phys.} \textbf{2015}, \emph{17},
  21866--21879\relax
\mciteBstWouldAddEndPuncttrue
\mciteSetBstMidEndSepPunct{\mcitedefaultmidpunct}
{\mcitedefaultendpunct}{\mcitedefaultseppunct}\relax
\EndOfBibitem
\bibitem[Villaume \latin{et~al.}(2010)Villaume, Saue, and
  Norman]{Norman10_064105}
Villaume,~S.; Saue,~T.; Norman,~P. Linear Complex Polarization Propagator in a
  Four-Component Kohn--Sham Framework. \emph{J. Chem. Phys.} \textbf{2010},
  \emph{133}, 064105\relax
\mciteBstWouldAddEndPuncttrue
\mciteSetBstMidEndSepPunct{\mcitedefaultmidpunct}
{\mcitedefaultendpunct}{\mcitedefaultseppunct}\relax
\EndOfBibitem
\bibitem[Antoulas(2005)]{Antoulas2005}
Antoulas,~A.~C. In \emph{Advances in Design and Control 6}; {Ralph C. Smith
  North Carolina State University},, Ed.; Advances in Design and Control 6;
  SIAM, 2005; p 479\relax
\mciteBstWouldAddEndPuncttrue
\mciteSetBstMidEndSepPunct{\mcitedefaultmidpunct}
{\mcitedefaultendpunct}{\mcitedefaultseppunct}\relax
\EndOfBibitem
\bibitem[Benner \latin{et~al.}(1998)Benner, Mehrmann, and Xu]{beme1998}
Benner,~P.; Mehrmann,~V.; Xu,~H. {A numerically stable, structure preserving
  method for computing the eigenvalues of real Hamiltonian or symplectic
  pencils}. \emph{Numer. Math.} \textbf{1998}, \emph{78}, 329--358\relax
\mciteBstWouldAddEndPuncttrue
\mciteSetBstMidEndSepPunct{\mcitedefaultmidpunct}
{\mcitedefaultendpunct}{\mcitedefaultseppunct}\relax
\EndOfBibitem
\bibitem[Bai and Li(2012)Bai, and Li]{bali2012}
Bai,~Z.; Li,~R.-C. {Minimization principles for the linear response eigenvalue
  problem I: Theory}. \emph{SIAM J. Mat. Anal. Appl.} \textbf{2012}, \emph{33},
  1075--1100\relax
\mciteBstWouldAddEndPuncttrue
\mciteSetBstMidEndSepPunct{\mcitedefaultmidpunct}
{\mcitedefaultendpunct}{\mcitedefaultseppunct}\relax
\EndOfBibitem
\bibitem[Shao \latin{et~al.}(2016)Shao, da~Jornada, Yang, Deslippe, and
  Louie]{SJYDL2016}
Shao,~M.; da~Jornada,~F.~H.; Yang,~C.; Deslippe,~J.; Louie,~S.~G. Structure
  preserving parallel algorithms for solving the Bethe--Salpeter eigenvalue
  problem. \emph{Linear Algebra Appl.} \textbf{2016}, \emph{488},
  148--167\relax
\mciteBstWouldAddEndPuncttrue
\mciteSetBstMidEndSepPunct{\mcitedefaultmidpunct}
{\mcitedefaultendpunct}{\mcitedefaultseppunct}\relax
\EndOfBibitem
\bibitem[Weiss \latin{et~al.}(1993)Weiss, Ahlrichs, and
  H\"{a}ser]{Haser93_1262}
Weiss,~H.; Ahlrichs,~R.; H\"{a}ser,~M. A Direct Algorithm for
  Self‐Consistent‐Field Linear Response Theory and Application to C60:
  Excitation Energies, Oscillator Strengths, and Frequency‐Dependent
  Polarizabilities. \emph{J. Chem. Phys.} \textbf{1993}, \emph{99},
  1262--1270\relax
\mciteBstWouldAddEndPuncttrue
\mciteSetBstMidEndSepPunct{\mcitedefaultmidpunct}
{\mcitedefaultendpunct}{\mcitedefaultseppunct}\relax
\EndOfBibitem
\bibitem[Stratmann \latin{et~al.}(1998)Stratmann, Scuseria, and
  Frisch]{Frisch98_8218}
Stratmann,~R.~E.; Scuseria,~G.~E.; Frisch,~M.~J. An Efficient Implementation of
  Time-Dependent Density-Functional Theory for the Calculation of Excitation
  Energies of Large Molecules. \emph{J. Chem. Phys.} \textbf{1998}, \emph{109},
  8218--8224\relax
\mciteBstWouldAddEndPuncttrue
\mciteSetBstMidEndSepPunct{\mcitedefaultmidpunct}
{\mcitedefaultendpunct}{\mcitedefaultseppunct}\relax
\EndOfBibitem
\bibitem[Brabec \latin{et~al.}()Brabec, Lin, Shao, Govind, Yang, Saad, and
  Ng]{brabec_etal2015}
Brabec,~J.; Lin,~L.; Shao,~M.; Govind,~N.; Yang,~C.; Saad,~Y.; Ng,~E.~G.
  Efficient Algorithms for Estimating the Absorption Spectrum within Linear
  Response TDDFT. \emph{J. Chem. Theor. Comput.} \emph{11}, 5197--5208\relax
\mciteBstWouldAddEndPuncttrue
\mciteSetBstMidEndSepPunct{\mcitedefaultmidpunct}
{\mcitedefaultendpunct}{\mcitedefaultseppunct}\relax
\EndOfBibitem
\bibitem[Li \latin{et~al.}(2017)Li, Valeev, Williams-Young, Ding, Liu, Goings,
  Petrone, and Lestrange]{chronusq_beta}
Li,~X.; Valeev,~E.~F.; Williams-Young,~D.; Ding,~F.; Liu,~H.; Goings,~J.;
  Petrone,~A.; Lestrange,~P. Chronus Quantum, Beta Version. 2017;
  \url{http://www.chronusquantum.org}\relax
\mciteBstWouldAddEndPuncttrue
\mciteSetBstMidEndSepPunct{\mcitedefaultmidpunct}
{\mcitedefaultendpunct}{\mcitedefaultseppunct}\relax
\EndOfBibitem
\bibitem[Walker(1988)]{Walker88_152}
Walker,~H.~F. Implementation of the GMRES Method using Householder
  Transformations. \emph{{SIAM J. Sci. Stat. Comp.}} \textbf{1988}, \emph{9},
  152--163\relax
\mciteBstWouldAddEndPuncttrue
\mciteSetBstMidEndSepPunct{\mcitedefaultmidpunct}
{\mcitedefaultendpunct}{\mcitedefaultseppunct}\relax
\EndOfBibitem
\bibitem[Shao \latin{et~al.}(2016)Shao, Aktulga, Yang, Ng, Maris, and
  Vary]{shak2016}
Shao,~M.; Aktulga,~H.~M.; Yang,~C.; Ng,~E.~G.; Maris,~P.; Vary,~J.~P.
  Accelerating Nuclear Configuration Interaction Calculations through a
  Preconditioned Block Iterative Eigensolver. \emph{arXiv} \textbf{2016},
  \emph{abs/1609.01689}\relax
\mciteBstWouldAddEndPuncttrue
\mciteSetBstMidEndSepPunct{\mcitedefaultmidpunct}
{\mcitedefaultendpunct}{\mcitedefaultseppunct}\relax
\EndOfBibitem
\bibitem[McLachlan and Ball(1964)McLachlan, and Ball]{Ball64_844}
McLachlan,~A.; Ball,~M. Time-Dependent Hartree--Fock Theory for Molecules.
  \emph{Rev. Mod. Phys.} \textbf{1964}, \emph{36}, 844\relax
\mciteBstWouldAddEndPuncttrue
\mciteSetBstMidEndSepPunct{\mcitedefaultmidpunct}
{\mcitedefaultendpunct}{\mcitedefaultseppunct}\relax
\EndOfBibitem
\bibitem[Harris(1969)]{Harris69_3947}
Harris,~R.~A. Oscillator Strengths and Rotational Strengths in Hartree--Fock
  Theory. \emph{J. Chem. Phys.} \textbf{1969}, \emph{50}, 3947--3951\relax
\mciteBstWouldAddEndPuncttrue
\mciteSetBstMidEndSepPunct{\mcitedefaultmidpunct}
{\mcitedefaultendpunct}{\mcitedefaultseppunct}\relax
\EndOfBibitem
\bibitem[Yeager \latin{et~al.}(1975)Yeager, Nascimento, and
  McKoy]{McKoy75_1168}
Yeager,~D.; Nascimento,~M.; McKoy,~V. Some Applications of
  Excited-State-Excited-State Transition Densities. \emph{Phys. Rev. A}
  \textbf{1975}, \emph{11}, 1168\relax
\mciteBstWouldAddEndPuncttrue
\mciteSetBstMidEndSepPunct{\mcitedefaultmidpunct}
{\mcitedefaultendpunct}{\mcitedefaultseppunct}\relax
\EndOfBibitem
\bibitem[Shao and Yang(2016)Shao, and Yang]{bsepack}
Shao,~M.; Yang,~C. BSEPACK User's Guide. 2016;
  \url{https://sites.google.com/a/lbl.gov/bsepack/}\relax
\mciteBstWouldAddEndPuncttrue
\mciteSetBstMidEndSepPunct{\mcitedefaultmidpunct}
{\mcitedefaultendpunct}{\mcitedefaultseppunct}\relax
\EndOfBibitem
\bibitem[St{\"o}hr(2013)]{Stohr_book}
St{\"o}hr,~J. \emph{NEXAFS Spectroscopy}; Springer Science \& Business Media,
  2013; Vol.~25\relax
\mciteBstWouldAddEndPuncttrue
\mciteSetBstMidEndSepPunct{\mcitedefaultmidpunct}
{\mcitedefaultendpunct}{\mcitedefaultseppunct}\relax
\EndOfBibitem
\end{mcitethebibliography}

\end{document}